\documentclass[a4paper,11pt]{article}
\usepackage[T1]{fontenc}
\usepackage{lmodern}

\usepackage{amsmath,amssymb}
\usepackage[page]{appendix}
\usepackage{array}
\usepackage{authblk}
\usepackage{bbm}
\usepackage{bm}
\usepackage{cite}
\usepackage{comment}
\usepackage{enumitem}
\usepackage{geometry}
\usepackage{graphicx}
\usepackage[scr=boondox]{mathalpha}
\usepackage{mathtools}
\usepackage{physics}
\usepackage{slashed}
\usepackage{subcaption}
\usepackage{subfiles}
\usepackage{xcolor}
\usepackage{tikz}
\usepackage{tikz}
\usetikzlibrary{arrows,shapes}
\usetikzlibrary{trees,patterns}
\usetikzlibrary{matrix,arrows} 				
\usetikzlibrary{positioning}				  
\usetikzlibrary{calc,through}				  
\usetikzlibrary{decorations.pathreplacing}  
\usepackage{pgffor}							

\usetikzlibrary{decorations.pathmorphing}	
\usetikzlibrary{decorations.markings}
\tikzset{
>=stealth', 
vector/.style={decorate, decoration={snake}, draw},
provector/.style={decorate, decoration={snake,amplitude=2.5pt}, draw},
antivector/.style={decorate, decoration={snake,amplitude=-2.5pt}, draw},
bigvector/.style={decorate, decoration={snake,amplitude=4pt}, draw},
fermion/.style={draw=black, postaction={decorate}, 
	decoration={markings,mark=at position .55 with {\arrow[draw=black]{>}}}},
fermionbar/.style={draw=black, postaction={decorate},
    decoration={markings,mark=at position .55 with {\arrow[draw=black]{<}}}},
fermionnoarrow/.style={draw=black},
doublefermion/.style={draw=black,double, postaction={decorate},
	decoration={markings,mark=at position .57 with {\arrow[draw=black]{>}}}},
doublefermionbar/.style={draw=black,double, postaction={decorate},
	decoration={markings,mark=at position .57 with {\arrow[draw=black]{<}}}},
doublefermionnoarrow/.style={draw=black,double},
gluon/.style={decorate, draw=black,
    decoration={coil,amplitude=4pt, segment length=5pt}},
scalar/.style={dashed,draw=black, postaction={decorate},
	decoration={markings,mark=at position .55 with {\arrow[draw=black]{>}}}},
scalarbar/.style={dashed,draw=black, postaction={decorate},
    decoration={markings,mark=at position .55 with {\arrow[draw=black]{<}}}},
scalarnoarrow/.style={dashed,draw=black},
momentum/.style={draw=black, postaction={decorate},
    decoration={markings,mark=at position 1 with {\arrow[draw=black]{>}}}},
antimomentum/.style={draw=black, postaction={decorate},
    decoration={markings,mark=at position 0.1 with {\arrow[draw=black]{<}}}}
}

\tikzstyle{block} = [draw, rectangle, minimum height=3em, minimum width=6em]

\usepackage[colorlinks=true,breaklinks=true]{hyperref}
\hypersetup{
allcolors=[rgb]{0.0 0.0 0.6},
linkcolor=[rgb]{0.0 0.0 1}
}

\newcommand{\aB}{\alpha_{\scriptscriptstyle\mathcal{B}}}

\newcommand{\aS}{\alpha_{\scriptscriptstyle\mathcal{S}}}

\newcommand{\arho}{\alpha_{\varrho}}
\newcommand{\aV}{\alpha_{\scriptscriptstyle{V}}}
\newcommand{\average}[1]{\left\langle #1 \right\rangle}

\newcommand{\BSF}{{\scriptscriptstyle{\rm BSF}}}

\newcommand{\DM}{{\scriptscriptstyle{\rm DM}}}
\newcommand{\elas}{{\rm elas}}
\newcommand{\im}{\mathbbm{i}}
\newcommand{\inel}{{\rm inel}}

\newcommand{\mrho}{m_{\varrho}}
\newcommand{\mT}{m_{\scriptscriptstyle{T}}}

\newcommand{\mX}{m_{\scriptscriptstyle{X}}}

\newcommand{\qX}{q_{\scriptscriptstyle{X}}}
\newcommand{\qrho}{q_{\varrho}}
\newcommand{\reg}{{\rm reg}}
\newcommand{\SIDM}{{\scriptscriptstyle{\rm SIDM}}}
\newcommand{\sym}{\mathscr{C}}

\newcommand{\twoPI}{{\scriptscriptstyle{\rm 2PI}}}
\newcommand{\unreg}{{\rm unreg}}
\newcommand{\vesc}{v_{\rm esc}}
\newcommand{\vrel}{v_{\rm rel}}
\newcommand{\y}{\mathrm{y}}
\newcommand{\zetaB}{\zeta_{\scriptscriptstyle{\cal B}}}
\newcommand{\zetaS}{\zeta_{\scriptscriptstyle{\cal S}}}
\newcommand{\zetan}{\zeta_n}

\newcommand{\calM}{{\scriptscriptstyle{\cal M}}}
\newcommand{\T}{{\scriptscriptstyle{\rm T}}}
\newcommand{\U}{{\scriptscriptstyle{\rm U}}}

\newcommand{\V}{{\scriptscriptstyle{\rm V}}}
\newcommand{\XX}{{\scriptscriptstyle{XX}}}
\newcommand{\XXdagger}{{\scriptscriptstyle{XX^\dagger}}}
\newcommand{\XdaggerXdagger}{{\scriptscriptstyle{X^\dagger X^\dagger}}}
\newcommand{\Yuk}{{\scriptscriptstyle{\rm Yuk}}}

\newcommand{\ID}{\mathbbm{1}}

\newcommand{\WW}{\mathbbm{W}}

\colorlet{lightblue}{blue!30}
\colorlet{lightred}{red!30}

\AtBeginEnvironment{appendices}{\crefalias{section}{appendix}}
\numberwithin{equation}{section}

\title{\bf 
Self-interacting dark matter from radiative bound-state-formation feedback
}
\author[1,2]{\Large
Marcos M. Flores\footnote{marcos.flores@fys.uio.no}}  
\author[1]{Kalliopi Petraki\footnote{kalliopi.petraki@phys.ens.psl.eu}}
\author[1]{Thomas Pignard\footnote{thomas.pignard@phys.ens.psl.eu}}
\affil[1]{\large\it
Laboratoire de Physique de l'École Normale Supérieure, 
ENS, 
Université PSL, 
CNRS, 
Sorbonne Université, 
Université Paris Cité,
Paris, F-75005, France}
\affil[2]{\large\it 
Department of Physics, University of Oslo, Box 1048, N-0316 Oslo, Norway }

\usepackage[capitalise]{cleveref}
\crefformat{pluralequation}{#2{\color{black} Eqs.~(}#1{\color{black} )}#3}
\Crefformat{pluralequation}
{#2{\color{black} Equations~(}#1{\color{black} )}#3}

\begin{document}

\maketitle
\begin{abstract}
Observations of galactic structure motivate sizable dark-matter self-interactions at low velocities, naturally realized through light force mediators. Such mediators can also make inelastic processes very efficient. Unitarity then requires their absorptive contributions to the two-particle self-energy kernel to be resummed, which in turn feeds back into both the elastic and inelastic scattering amplitudes. We study the feedback of radiative bound-state formation on elastic scattering and its implications for self-interacting dark matter. We focus on capture processes through emission of a charged scalar; such transitions are particularly efficient because the scattering and bound states are governed by different effective potentials, a feature also shared by radiative capture through emission of a non-Abelian gauge boson. We show that their feedback can generate sizable elastic scattering relevant for halo dynamics, even when the baseline elastic amplitude vanishes. Notably, the generated elastic cross section can grow toward low velocities, resembling long-range scattering. A companion paper examines the feedback from annihilation processes.

\end{abstract}

\clearpage
\noindent\rule{\textwidth}{0.4pt}
\tableofcontents
\noindent\rule{\textwidth}{0.4pt}
\clearpage

\section{Introduction}
\label{Sec:Intro}

Galactic structure observations motivate dark matter (DM) self-interactions that are sizable at the low characteristic velocities of dwarf galaxies and subhalos, while remaining sufficiently weak at the larger velocities of galaxy clusters~\cite{Spergel:1999mh,Tulin:2017ara,Bullock:2017xww,Markevitch:2003at,Randall:2008ppe,Dutra:2024qac,Enzi:2024ygw,Vegetti:2026mmx,Gilman:2026wul}. Such a velocity dependence arises naturally when DM interacts through light force mediators~\cite{Feng:2009hw,Loeb:2010gj,Tulin:2012wi,Tulin:2013teo}. If the mediator mass is comparable to or smaller than the characteristic momentum transfer, the interaction becomes effectively long-ranged and must generally be treated nonperturbatively. The resulting Coulomb- or Yukawa-like dynamics can produce large low-velocity self-scattering cross sections while remaining sufficiently weak at higher velocities, providing a particle-physics realization of self-interacting dark matter (SIDM).

The presence of light force mediators can also make inelastic processes important. Long-range interactions can strongly enhance annihilation and radiative processes such as bound-state formation (BSF), allowing their partial-wave cross sections to approach the limits imposed by unitarity~\cite{vonHarling:2014kha,Petraki:2015hla,Petraki:2016cnz,Baldes:2017gzw,Flores:2024sfy}. In this regime, elastic and inelastic scattering cannot be treated independently. The optical theorem imposes coupled constraints on their amplitudes in each partial wave~\cite{Hui:2001wy,Flores:2024sfy,Flores:2025uoh}, such that a calculation that resums only the purely elastic interaction kernel is in general insufficient once inelasticity becomes sizable~\cite{Blum:2016nrz,Flores:2024sfy,Flores:2025uoh}. The absorptive contributions generated by open inelastic channels must also be incorporated into the two-particle self-energy kernel of the incoming state. Their resummation feeds back into both elastic and inelastic scattering and restores the coupled unitarity constraints. Importantly, this feedback need not merely suppress elasticity: when the baseline elastic amplitude is small or absent, inelastic transitions through intermediate states can instead generate a nonzero elastic cross section~\cite{Flores:2024sfy,Flores:2025uoh}.

Radiative BSF provides a particularly striking realization of these dynamics when the emitted boson carries charge, as can occur in Abelian and non-Abelian theories through the emission of a charged scalar~\cite{Oncala:2019yvj,Oncala:2021tkz,Oncala:2021swy} or a non-Abelian gauge boson~\cite{Harz:2018csl,Binder:2023ckj}. In this case, the incoming scattering state and outgoing bound state are governed by different interaction potentials, allowing very rapid capture transitions. Summing capture over the tower of excited bound levels leads to an especially strong low-velocity enhancement: the standard Coulombic calculation exceeds the global inelastic partial-wave unitarity limit at sufficiently low velocities, even for arbitrarily small nonzero couplings~\cite{Beneke:2024nxh,Flores:2026yay}. This apparent unitarity violation signals the breakdown of treating BSF as an isolated process. A consistent resummation of the BSF-induced contribution to the incoming-state kernel is then essential~\cite{Flores:2026yay} and, through the optical theorem, modifies the elastic scattering amplitude as well~\cite{Flores:2024sfy,Flores:2025uoh}. A complementary effective-field-theory approach to the unitarization of radiative capture and the associated generation of elastic scattering has recently been developed~\cite{DeRos:2026lhi}.

In this work, we investigate the feedback of BSF via charged-scalar emission on elastic scattering. We consider scalar DM $X$ and a light scalar mediator $\varrho$, singly and doubly charged under a dark $U(1)$ symmetry, respectively. We work in the limit in which this symmetry is global, such that no baseline long-range potential acts between same-charge particles, while particle-antiparticle pairs can bind into bound states ${\cal B}(XX^\dagger)$ and same-charge particles can undergo the capture process $XX\to {\cal B}(XX^\dagger)+\varrho$~\cite{Oncala:2019yvj}.  While transitions proceeding via emission of a vector or neutral scalar boson are suppressed by dipole and quadrupole moments, respectively~\cite{Pospelov:2008jd,March-Russell:2008klu,Wise:2014jva,Petraki:2016cnz,Oncala:2018bvl}, the non-orthogonality of the scattering and bound-state wavefunctions in the charged-scalar process permits an unsuppressed monopole transition, making capture particularly efficient~\cite{Oncala:2019yvj,Ko:2019wxq}. This provides an additional enhancement on top of that arising from the different interaction potentials governing the incoming and outgoing states. On the other hand, the finite mass of the emitted scalar provides an important countervailing effect by imposing a kinematic cutoff on the accessible bound levels. As the velocity decreases, progressively fewer excited states remain kinematically accessible, eventually curtailing the growth that would otherwise occur for a massless emitted boson. Moreover, the unitarized result is sensitive to the nontrivial analytic structure of the underlying BSF amplitudes in the complex momentum plane~\cite{Flores:2026yay}.

We find that BSF feedback can generate sizable elastic self-scattering even when no baseline elastic potential acts in the incoming channel. The resulting elastic cross section can reach values relevant for halo dynamics in regions where the baseline BSF rate becomes very large. Importantly, it can grow toward low velocities, providing a novel realization of SIDM that can satisfy the elastic self-interaction benchmarks.

The phenomenological implications differ qualitatively between symmetric and asymmetric DM. For symmetric DM, the BSF-induced $XX$ scattering supplements the conventional $XX^\dagger$ Yukawa scattering. In the parameter space we examine, however, the latter remains dominant once the phase-space suppression of the feedback contribution due to the finite mediator mass and the effects associated with the analytic structure of the BSF amplitudes are included. An exception occurs near Yukawa antiresonances, where the BSF-induced scattering can dominate; in the parameter space explored, however, these regions do not open additional viable SIDM parameter space. For asymmetric DM, by contrast, only the same-charge population survives, and the baseline long-range elastic interaction vanishes in the limit considered here. BSF feedback can nevertheless generate elastic self-scattering that grows toward low velocities, thereby opening a qualitatively different realization of SIDM. These results complement our companion study of annihilation feedback, where the irreducible annihilation interaction is short-distance and affects elastic scattering appreciably mainly near Sommerfeld resonances and zeros of the baseline elastic amplitude~\cite{Petraki:2026SIDMann}.

The remainder of this paper is organized as follows. In \cref{Sec:BSFbaseline}, we introduce the dark-sector model and present the baseline BSF amplitudes and cross sections, including the sum over bound levels and the kinematic restrictions associated with the finite mass of the emitted scalar. In \cref{Sec:Feedback}, we unitarize the BSF amplitudes and investigate their feedback on elastic scattering, paying particular attention to the generation of elastic scattering in a channel with no baseline elastic potential and to the analytic structure of the capture amplitudes. In \cref{Sec:SIDM}, we assess the implications for the SIDM parameter space for symmetric and asymmetric DM. We conclude in \cref{Sec:Conclusions}.

\clearpage
\section{Bound-state formation}
\label{Sec:BSFbaseline}

\subsection{Model}
\label{Sec:BSFbaseline_Model}

We consider the scalar dark sector introduced in Ref.~\cite{Oncala:2019yvj} and studied in our companion work~\cite{Petraki:2026SIDMann}, specializing from the outset to the global-$U(1)_D$ limit. The gauged theory and possible realizations of this limit are discussed in Refs.~\cite{Petraki:2026SIDMann,Oncala:2019yvj}.

The DM field $X$ and the mediator $\varrho$ are complex scalars with charges $\qX=1$ and $\qrho=2$ under the  $U(1)_D$ symmetry, respectively~\cite{Oncala:2019yvj}. This charge assignment allows the trilinear interaction $X^2\varrho^\dagger$. The terms relevant for the dynamics considered here are
\begin{align}
{\cal L}
&
=|\partial_\mu X|^2 - \mX^2|X|^2
+|\partial_\mu\varrho|^2
-\mrho^2|\varrho|^2
-\dfrac{\y \mX}{2}
\qty[X^2\varrho^\dagger+(X^\dagger)^2\varrho]
+{\cal L}_{\rm quartic}.
\label{eq:Lagrangian}
\end{align}
We take $\mrho \ll \mX$ and define
\begin{align}
\arho\equiv\dfrac{\y^2}{16\pi}.
\label{eq:alphaRho}
\end{align}
The quartic interactions are collected in
\begin{align}
{\cal L}_{\rm quartic} =
-\dfrac{\lambda_X}{4}|X|^4
-\dfrac{\lambda_\varrho}{4}|\varrho|^4
-\lambda_{X\varrho}|X|^2|\varrho|^2.
\label{eq:Lquartic}
\end{align}
We assume that the scalar-potential parameters are such that $X=\varrho=0$ is a stable or cosmologically long-lived vacuum.

\subsubsection*{Long-range potentials and bound states}

After partial-wave projection, the $u$-channel $\varrho$ exchange generates a Yukawa potential between $X$ and $X^\dagger$~\cite{Oncala:2019yvj}, while no long-range interaction exists between same-charge particles, 
\begin{align}
V_{XX^\dagger,\ell}(r) =
-(-1)^\ell \, \arho \,
\dfrac{e^{-\mrho r}}{r}
\qquad
\text{and}
\qquad
V_{XX}(r)
=V_{X^\dagger X^\dagger}(r)
=0 .
\label{eq:Potentials}
\end{align}
Quartic interactions and virtual $s$-channel $\varrho$ exchange also generate contact-like contributions to $XX$ elastic scattering. Unlike BSF-induced scattering, these contributions are not enhanced at low velocities and can become subdominant in the light-mediator regime considered here, as follows from the results of \cref{Sec:Feedback_VelocityDepend}. We neglect them to isolate the effect of BSF feedback on elastic scattering.

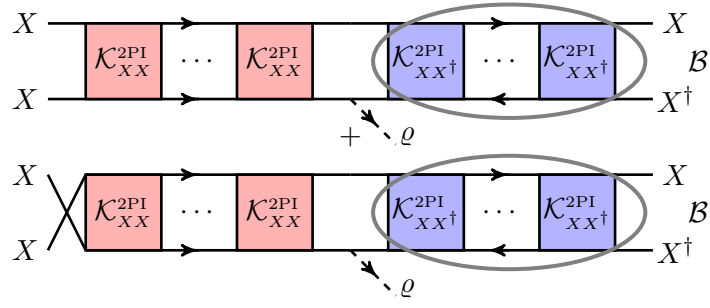
\begin{figure}[t!]
\centering
\begin{tikzpicture}[line width=1pt, scale=1]
\begin{scope}[shift={(0,0.5)}]
\node at (-4.3,1){$X$};
\node at (-4.3,0){$X$};
\draw[fermion]	(-2.1,1) -- (-2,1);
\draw[fermion]	(-2.1,0) -- (-2,0);
\draw (-4,1) -- (0,1);
\draw (-4,0) -- (0,0);
%
\draw[fill=lightred,shift={(-3,0.5)}] (-0.5,-0.5) rectangle (0.5,0.5);
\draw[fill=lightred,shift={(-1,0.5)}] (-0.5,-0.5) rectangle (0.5,0.5);
\node at (-3,0.5){${\cal K}_{\scriptscriptstyle{XX}}^{\twoPI}$};
\node at (-1,0.5){${\cal K}_{\scriptscriptstyle{XX}}^{\twoPI}$};
\node at (-2,0.5){$\cdots$};
\draw[scalar]	(0,0) -- (0.6,-0.6);
\node at (0.75,-0.5){$\varrho$};
\node at (4.3,1){$X$};
\node at (4.3,0){$X^{\dagger}$};
\draw[fermion]		(2,1) -- (2.1,1);
\draw[fermionbar]	(2,0) -- (2.1,0);
\draw (0,1) -- (4,1);
\draw (0,0) -- (4,0);
%
\draw[fill=lightblue,shift={(3,0.5)}] (-0.5,-0.5) rectangle (0.5,0.5);
\draw[fill=lightblue,shift={(1,0.5)}] (-0.5,-0.5) rectangle (0.5,0.5);
\node at (3,0.5){${\cal K}_{\scriptscriptstyle{XX^{\dagger}}}^{\twoPI}$};
\node at (1,0.5){${\cal K}_{\scriptscriptstyle{XX^{\dagger}}}^{\twoPI}$};
\node at (2,0.5){$\cdots$};
\draw[fill=none,gray,line width=1.5pt] (2.1,0.5) ellipse (1.8 and 0.75);
\node at (4.6,0.5){${\cal B}$};
\end{scope}
%
\node at (0,0){$+$};
%
\begin{scope}[shift={(0,-1.5)}]
\node at (-4.3,1){$X$};
\node at (-4.3,0){$X$};
\draw[fermion]	(-2.1,1) -- (-2,1);
\draw[fermion]	(-2.1,0) -- (-2,0);
\draw (-4,0) -- (-3.5,1);\draw (-3.5,1) -- (0,1);
\draw (-4,1) -- (-3.5,0);\draw (-3.5,0) -- (0,0);
%
\draw[fill=lightred,shift={(-3,0.5)}] (-0.5,-0.5) rectangle (0.5,0.5);
\draw[fill=lightred,shift={(-1,0.5)}] (-0.5,-0.5) rectangle (0.5,0.5);
\node at (-3,0.5){${\cal K}_{\scriptscriptstyle{XX}}^{\twoPI}$};
\node at (-1,0.5){${\cal K}_{\scriptscriptstyle{XX}}^{\twoPI}$};
\node at (-2,0.5){$\cdots$};
\draw[scalar]	(0,0) -- (0.6,-0.6);
\node at (0.75,-0.5){$\varrho$};
\node at (4.3,1){$X$};
\node at (4.3,0){$X^{\dagger}$};
\draw[fermion]		(2,1) -- (2.1,1);
\draw[fermionbar]	(2,0) -- (2.1,0);
\draw (0,1) -- (4,1);
\draw (0,0) -- (4,0);
%
\draw[fill=lightblue,shift={(3,0.5)}] (-0.5,-0.5) rectangle (0.5,0.5);
\draw[fill=lightblue,shift={(1,0.5)}] (-0.5,-0.5) rectangle (0.5,0.5);
\node at (3,0.5){${\cal K}_{\scriptscriptstyle{XX^{\dagger}}}^{\twoPI}$};
\node at (1,0.5){${\cal K}_{\scriptscriptstyle{XX^{\dagger}}}^{\twoPI}$};
\node at (2,0.5){$\cdots$};
\draw[fill=none,gray,line width=1.5pt] (2.1,0.5) ellipse (1.8 and 0.75);
\node at (4.6,0.5){${\cal B}$};
\end{scope}
\end{tikzpicture}
\caption{Capture into a bound state via charged-scalar emission, $XX\to \mathcal{B}(XX^\dagger)+\varrho$. In the model considered here, the long-range $XX^\dagger$ kernel is generated by $u$-channel $\varrho$ exchange, whereas no baseline long-range kernel is present in the incoming $XX$ channel. Nevertheless, BSF generates an absorptive contribution to ${\cal K}_\XX^\twoPI$, illustrated in \cref{fig:BSF_Kernel_BSFsquared}, which, upon resummation, modifies the BSF cross sections and induces elastic $XX$ scattering.}
\label{fig:BSF_Transition}
\end{figure}

The $XX^\dagger$ potential is attractive for even partial waves and repulsive for odd ones. Therefore, it supports $XX^\dagger$ bound states only in the even partial waves, provided that $\varrho$ is sufficiently light. Despite the absence of a long-range interaction, same-charge pairs can form $XX^\dagger$ bound states through $\varrho$ emission,
\begin{align}
X+X
\to
{\cal B}_{n\ell}(XX^\dagger)+\varrho,
\label{eq:BSF_Process}
\end{align}
and the conjugate process, as shown in \cref{fig:BSF_Transition}. Here, $n\geqslant 1+\ell$ stands for the principal quantum number characterizing bound states in a central potential. We define
\begin{align}
\mT = 2\mX, 
\qquad
\mu= \mX/2,
\qquad
\kappa \equiv \mu \arho,
\qquad
\kappa_n \equiv \kappa/n,
\qquad
\vb{k} = \mu \vb{v}_{\rm rel},
\label{eq:BSF_scales}
\end{align}
where $\mT$ and $\mu$ are the total and reduced masses of the interacting particles in the scattering or bound states, $\kappa$ is the Bohr momentum, $\kappa_n$ is the binding momentum of the $n$th Coulombic bound level, and $\vb{k}$ is the momentum of one of the scattering-state particles in the center-of-momentum (CM) frame, with $\vb{v}_{\rm rel}$ being the relative velocity; we will also use the magnitudes $k=|\vb{k}|=\mu\vrel$ and $\vrel = |\vb{v}_{\rm rel}|$. The scattering-state kinetic energy in the CM frame and the Coulombic binding energies are
\begin{align}
{\cal E}_{\vb{k}} = \dfrac{\vb{k}^2}{2\mu}
\quad \text{and} \quad
{\cal E}_n = -\frac{\kappa_n^2}{2\mu},
\label{eq:BSF_KineticAndBindingEnergies}
\end{align}
respectively. In the Coulomb approximation of the bound-state potential, the processes \eqref{eq:BSF_Process} are kinematically allowed if $\mrho < \omega_{n}(k)$, where $\omega_{n}$ is the energy radiated in the capture process~\cite{Petraki:2015hla,Petraki:2016cnz} 
\begin{align}  
\omega_{n} (k) \equiv
{\cal E}_{\vb{k}} - {\cal E}_n = 
\dfrac{\mu}{2} \qty(\vrel^2 + \dfrac{\arho^2}{n^2}).
\label{eq:BSF_EnergyDissipated}
\end{align}
Capture processes are typically significant at low velocities, when ${\cal E}_{\vb{k}}\lesssim |{\cal E}_n|$, so the kinematic condition becomes $\mrho\lesssim |{\cal E}_n|$. For perturbative $\arho$, this implies
\begin{align}
\mrho
\lesssim \dfrac{\kappa^2}{2n^2\mu}
\ll \dfrac{\kappa}{n^2}
\leqslant \dfrac{\kappa}{n},
\label{eq:LengthScales_BoundStates}
\end{align}
such that the interaction range, $\mrho^{-1}$, exceeds the characteristic bound-state length scales. We therefore use Coulombic bound-state wavefunctions in the BSF amplitudes, while retaining the finite mediator mass in the emission kinematics. We examine the validity of this approximation further in \cref{Sec:BSFbaseline_CoulombApproximation}.

Next, we present the baseline amplitudes for the BSF processes \eqref{eq:BSF_Process}. For convenience, we define the dimensionless parameters
\begin{subequations}
\label{eq:NoDimParameters}
\label{eqs:NoDimParameters}
\begin{gather}
\zeta \equiv \dfrac{\kappa}{k} = \dfrac{\arho}{\vrel},
\qquad
\zetan \equiv \dfrac{\kappa_n}{k} = \dfrac{\arho}{n\, \vrel},
\qquad
\xi \equiv \dfrac{\kappa}{\mrho} = \dfrac{\mu \arho}{\mrho},
\\[2ex]
\gamma \equiv \dfrac{\mrho}{|{\cal E}_{n=1}|} 
=\dfrac{2\mrho}{\mu \arho^2} 
= \dfrac{2}{\arho \, \xi}
.
\end{gather}
\end{subequations}

\subsection{Unregulated amplitudes and cross sections}
\label{Sec:BSFbaseline_Amplitude}

The baseline (unregulated) amplitudes for the BSF processes \eqref{eq:BSF_Process}, computed using the long-range potentials \eqref{eq:Potentials}, while neglecting BSF feedback, are~\cite{Oncala:2019yvj,Flores:2026yay}
\begin{align}
{\cal M}_{n\ell,\unreg}^{\BSF} (k) 
=  (\sym_\ell^{\XX})^{1/2}
\  \arho^{1/2}
\, \dfrac{\mT\sqrt{2\mu}}{2\kappa_n^{3/2}} 
\, r_{n\ell}^{\calM} (\zeta),
\label{eq:AmplitudeBSF_unreg}
\end{align}
where $\sym_\ell^\XX$ is the symmetry factor for the identical particles in the incoming $XX$ state,
\begin{align}
\sym_\ell^{\XX}=2\delta_{\ell,\rm even}.
\label{eq:SymmetryFactor_XX}
\end{align}
The monopole transition obeys the selection rule $\Delta\ell=0$, so that the orbital angular momentum of the final bound state coincides with that of the incoming scattering state. The momentum dependence of the amplitudes is encapsulated in the functions
\begin{align}
r_{n\ell}^{\calM} (\zeta) 
&= \im^\ell(-1)^{n-\ell} 
\dfrac{2^{2\ell+3}\ell!}{(2\ell+1)!}
\qty[\frac{n(n+\ell)!}{(n-\ell-1)!}]^{1/2}
\nonumber\\
&\times 
\dfrac{\zeta_n^{\ell+4}}{(\zeta_n\mp \im)^{2\ell+4}} 
\qty(\frac{\zeta_n\mp\im}{\zeta_n\pm \im})^{n+1} 
\, {}_2F_1 \qty(1+\ell-n,1+\ell;2\ell+2;\mp \frac{4\im \zeta_n}{(\zeta_n\mp \im)^2}).
\label{eq:AmplitudeBSF_unreg_rnlM}
\end{align}
The two signs in \cref{eq:AmplitudeBSF_unreg_rnlM} arise from a Pfaff transformation of the hypergeometric function  ${}_2F_1$. The corresponding cross sections are~\cite{Oncala:2019yvj,Flores:2026yay} 
\begin{align}
\sigma_{n\ell,\unreg}^{\BSF} (k) 
=
\sym_\ell^{\XX} \, 
\dfrac{4\pi (2\ell+1)}{k^2}
\times \Phi_n(k)
\times \arho
\times r_{n\ell} (\zeta),
\label{eq:CrossSectionBSF}
\end{align}
where $\Phi_n(k)$ is the phase-space suppression factor,  
\begin{align}
\Phi_n(k) \equiv 
\sqrt{1-\dfrac{\mrho^2}{\omega_n^2(k)}}
= \sqrt{1-\dfrac{\gamma^2\zeta^4}{(1+\zetan^2)^2}},
\label{eq:CrossSectionBSF_PhaseSpaceSuppr}
\end{align}
and 
\begin{align}
\label{eq:CrossSectionBSF_rnl}
&r_{n\ell} (\zeta) \equiv
\dfrac{1+\zetan^2}{\zetan^3} \times
|r_{n\ell}^{\calM} (\zeta)|^2
\\
&= \! \qty[\dfrac{2^{2\ell+3} \ell!}{(2\ell+1)!}]^2
\dfrac{n (n+\ell)!}{(n-\ell-1)!}
\dfrac{\zetan^{2\ell+5}}{(1+\zetan^2)^{2\ell+3}}
\abs{ 
{}_2F_1\qty[
\dfrac{-n+\ell+1}{2},
\dfrac{n+\ell+1}{2};
\ell+\dfrac{3}{2};
\dfrac{4\zetan^2}{(1+\zetan^2)^2}
]}^2,
\nonumber
\end{align}
where we used a quadratic transformation for ${}_2F_1$ of \cref{eq:AmplitudeBSF_unreg_rnlM}. The finite mediator mass can impose a kinematic cutoff on the levels that can form radiatively from free pairs at low velocities,
\begin{align}
n <
n_{\rm cut}^\BSF (k) \equiv 
\begin{cases}
+\infty, & \gamma\zeta^2 \leqslant 1,
\\
\dfrac{\zeta}{\sqrt{\gamma\zeta^2-1}}, 
\quad 
&\gamma\zeta^2 > 1.
\end{cases}
\label{eq:BSF_PhaseSpaceCutoff_ncut}
\end{align}
When $\mu \vrel^2/2 \geqslant \mrho$, or equivalently $\gamma\zeta^2 \leqslant 1$, the kinetic energy alone is sufficient to produce the emitted mediator. At lower velocities, the binding energy must contribute to reaching the emission threshold, restricting capture to sufficiently low-lying levels. In the $\zeta\to\infty$ limit, $n_{\rm cut}^{\BSF}\to\gamma^{-1/2}$. Since $n\geqslant1+\ell$, radiative capture into a state with angular momentum $\ell$ remains kinematically possible only if $\gamma<(1+\ell)^{-2}$.

\subsection{Excited levels and unitarity violation}
\label{Sec:BSFbaseline_ExcitedLevels}

\subsubsection*{Partial-wave unitarity bounds}

Unitarity sets coupled constraints on partial-wave elastic and inelastic cross sections~\cite{Hui:2001wy,Flores:2024sfy}. We define the partial-wave unitarity cross section~\cite{Flores:2024sfy,Flores:2025uoh},
\begin{align}
\sigma_\ell^\U(k)
\equiv
\sym_\ell\,\dfrac{4\pi(2\ell+1)}{k^2},
\label{eq:sigmaUnitarity_Elastic}
\end{align}
where $\sym_\ell=1$ for distinguishable particles and $\sym_\ell=2$ for surviving modes of identical particles; modes forbidden by exchange symmetry are omitted. We define the elastic and inclusive inelastic cross sections normalized to $\sigma_\ell^\U$,
\begin{align}
x_\ell^\elas
\equiv
\dfrac{\sigma_\ell^\elas}{\sigma_\ell^\U},
\qquad
y_\ell
\equiv
\dfrac{\sigma_\ell^\inel}{\sigma_\ell^\U}.
\label{eq:NormalizedCrossSections_def}
\end{align}
The unitarity constraints read~\cite{Flores:2024sfy, Flores:2025uoh},
\begin{gather}
\qty(x_\ell^\elas+y_\ell)^2
\leqslant
x_\ell^\elas
\qquad
\Rightarrow
\quad
x_\ell^\elas \leqslant 1,
\quad
y_\ell \leqslant 1/4.
\label{eq:UnitarityConstraint_xy}
\end{gather}
At the global inelastic maximum, $y_\ell=1/4$, the elastic cross section is fixed to $x_\ell^\elas=1/4$.

In SIDM phenomenology, the transfer and viscosity cross sections are also commonly employed to characterize momentum and heat transport within halos. Unlike the elastic cross section, their partial-wave contributions involve interference between different angular modes. Unitarity implies corresponding term-by-term upper bounds~\cite{Petraki:2026SIDMann},
\begin{subequations}
\label{eq:sigmaUnitarity_TV}
\label[pluralequation]{eqs:sigmaUnitarity_TV}
\begin{align}
x_\ell^\T
&\equiv
\dfrac{\sigma_\ell^\T}{\sigma_\ell^{\U,\T}}
\leqslant 1,
\qquad
\sigma_\ell^{\U,\T}(k)
\equiv
\dfrac{4\pi}{k^2}(\ell+1), 
\qquad
\text{distinguishable particles only,}
\label{eq:sigmaUnitarity_Transfer}
\\
x_\ell^\V
&\equiv
\dfrac{\sigma_\ell^\V}{\sigma_\ell^{\U,\V}}
\leqslant 1,
\qquad
\sigma_\ell^{\U,\V}(k)
\equiv
\sym_\ell
\dfrac{4\pi}{k^2}
\dfrac{\qty(\ell+1)\qty(\ell+2)}{2\ell+3}.
\label{eq:sigmaUnitarity_Viscosity}
\end{align}
\end{subequations}
For identical particles, the transfer cross section coincides with the elastic cross section, $\sigma^\T=\sigma^\elas$, so no separate transfer bound is required.

\subsubsection*{Bound-state tower and low-velocity growth}

To obtain the inclusive inelastic cross section for a given partial wave, we must sum capture over the tower of bound levels accessible from the same partial wave,
\begin{align}
\sigma_{\ell,\unreg}^\BSF (k) 
\equiv
\sum_{n = 1+\ell}^{n_{\rm cut}^\BSF(k)}    
\sigma_{n\ell,\unreg}^\BSF (k)
= \sigma_\ell^\U (k) \, \arho \, r_\ell (\zeta,\gamma) ,
\label{eq:BSF_sigmaell}
\end{align}
where\footnote{
Our notation differs slightly from that of Ref.~\cite{Flores:2026yay}, where the functions $r_{n\ell}(\zetaS,\zetaB)$ and $r_\ell(\zetaS,\zetaB)$ were defined for massless emission and general Coulomb couplings $\aS$ and $\aB$ in the scattering and bound states, respectively, with $\zetaS\equiv\aS/\vrel$ and $\zetaB\equiv\aB/\vrel$. Here we take $\aS=0$ and $\zeta=\zetaB$, but allow for a finite mediator mass.
Our $r_{n\ell}(\zeta)$ corresponds to $r_{n\ell}(0,\zeta)$ of Ref.~\cite{Flores:2026yay}, while our $r_\ell(\zeta,\gamma)$ reduces to $r_\ell(0,\zeta)$ of that reference only for $\gamma=0$.
}
\begin{align}
r_\ell (\zeta,\gamma) \equiv
\sum_{n = 1+\ell}^{n_{\rm cut}^\BSF(k)}   
\Phi_n(k) \, 
r_{n\ell} (\zeta) .
\label{eq:BSF_rell_def}
\end{align}
Consequently,
\begin{align}
y_{\ell,\unreg}^\BSF =
\arho\,r_\ell(\zeta,\gamma).
\label{eq:BSF_yUnreg}
\end{align}
For emission of a massless mediator, Ref.~\cite{Flores:2026yay} derived Kramers-like expressions for the sum over excited levels within each partial wave. At sufficiently low velocities,
\begin{align}
r_\ell(\zeta,0)
\approx
4\zeta
\qty[
\ln\qty(\dfrac{\zeta}{b_\ell})
+\dfrac{1}{2(1+\ell)}
],
\qquad
\zeta\gg(1+\ell)e,
\label{eq:BSF_Kramers}
\end{align}
where $b_\ell\simeq(1+\ell)/e$. Thus $r_\ell(\zeta,0)\sim4\zeta\ln\zeta$ asymptotically. This unbounded growth at large $\zeta$ implies that the unregulated partial-wave BSF cross sections exceed their global inelastic bounds at sufficiently low velocities for \emph{any} nonzero $\arho$~\cite{Flores:2026yay}.

\begin{figure}[t!]
\centering
\includegraphics[width=0.95\linewidth]{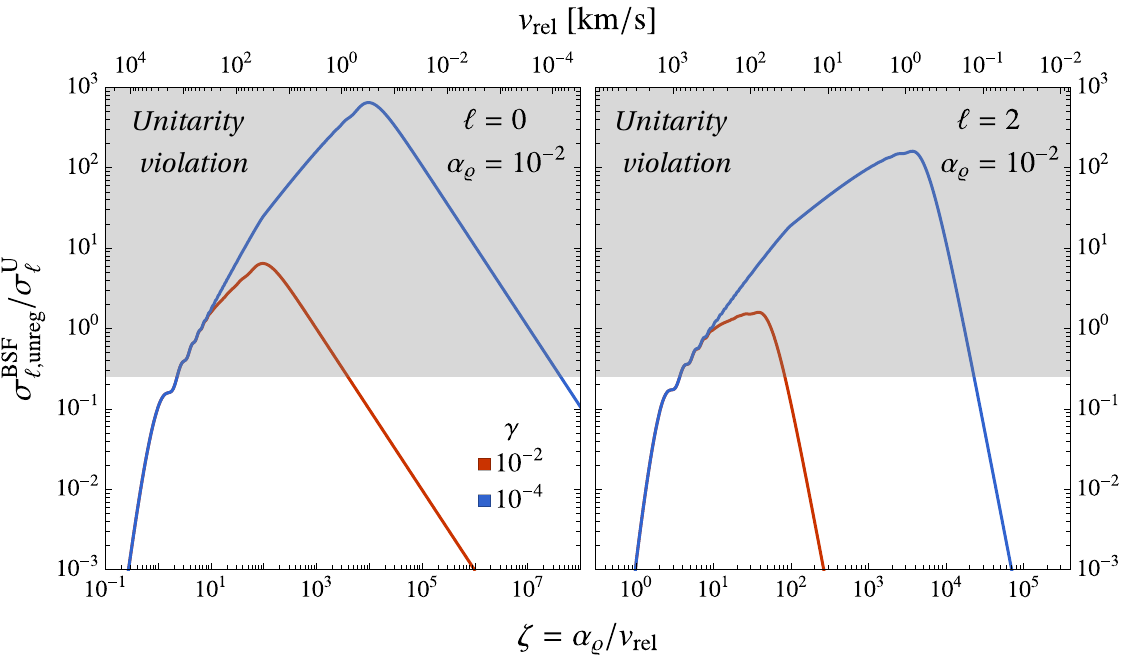}
\caption{
The unregulated inclusive BSF cross section $y_{\ell,\unreg}^{\BSF}$ grows at low velocities due to the increasingly important contribution of excited levels. For finite $\mrho$, the phase-space cutoff on the excited-state tower eventually curtails this growth.
We have used representative values of $\ell$, $\arho$, and $\gamma=2\mrho/(\mu\arho^2)$, as denoted in the legends.}
\label{fig:BSF_UniViolationAndRestoration}
\end{figure}

For finite $\mrho$, however, the phase-space cutoff progressively renders the highly excited levels responsible for this growth inaccessible. The unregulated BSF cross section can consequently exceed the global bound only over a finite range of velocities before being curtailed again at lower velocities, as illustrated in \cref{fig:BSF_UniViolationAndRestoration}. In \cref{App:BSFsums}, we derive analytic approximations for $r_\ell(\zeta,\gamma)$ at finite $\mrho$. Here we estimate when the cutoff curtails the low-velocity growth of the inclusive unregulated BSF cross section. In the limit $\mrho\to0$, Ref.~\cite{Flores:2026yay} estimated that the maximum principal quantum number required to reproduce the low-velocity growth of $y_{\ell,\unreg}^{\BSF}$, for a given $\ell$, to fractional precision $\mathtt{p}$ is
\begin{align}
n_{\rm growth}^{\BSF} (\mathtt{p})
&\approx
\dfrac{\zeta}{
\sqrt{\qty(\zeta/b_\ell)^{\mathtt{p}}-1}}
\approx
\dfrac{\zeta}{
\sqrt{\mathtt{p}\ln(\zeta/b_\ell)}},
\qquad
\zeta >b_\ell
\quad\text{and}\quad
\mathtt{p}\ll1.
\label{eq:BSF_ngrowth}
\end{align}
Retaining the Coulomb approximation for the capture amplitudes but taking into account the phase-space suppression, we estimate that a finite $\mrho$ affects this growth when
$n_{\rm cut}^{\BSF}<n_{\rm growth}^{\BSF}(\mathtt{p})$. Combining \cref{eq:BSF_PhaseSpaceCutoff_ncut,eq:BSF_ngrowth}, this condition becomes
\begin{align}
\zeta \gtrsim
\qty(\gamma b_\ell^{\mathtt{p}})^{-1/(2-\mathtt{p})}
\qquad 
\overset{\mathtt{p}\to 0}{\longrightarrow} 
\qquad
\zeta \gtrsim \gamma^{-1/2}
\quad
\text{or}
\quad
\vrel \lesssim \sqrt{2\mrho/\mu}
,
\label{eq:BSF_CutVsGrowth}
\end{align}
independently of $\ell$. 

\subsection{Validity of the Coulomb approximation with kinematic cutoff}
\label{Sec:BSFbaseline_CoulombApproximation}

The validity of the Coulomb approximation for the capture amplitudes can be assessed from the characteristic scales of the Coulombic bound states,
\begin{gather}
\average{|\vb{r}|} 
= \dfrac{1}{2\kappa} [3n^2-\ell(\ell+1)],
\quad
\langle\vb{r}^2\rangle 
= \dfrac{n^2}{2\kappa^2} [5n^2+1-3\ell(\ell+1)],
\quad
\langle\vb{p}^2\rangle 
= \dfrac{\kappa^2}{n^2}.
\label{eq:CoulombBoundStates_Averages}
\end{gather}
The Coulomb  approximation requires the characteristic bound-state length scale to be within the screening radius, $n^2/\kappa \ll \mrho^{-1}$ or equivalently $\xi \gg n^2$.

In our approximation, the levels that contribute significantly to the total capture rate are limited by the kinematic cutoff $n_{\rm cut}^\BSF$ and by the scale $n_{\rm growth}^\BSF$. We therefore expect the Coulomb approximation to be accurate for our purposes provided that
\begin{align}
\sqrt{\xi} \gg \min\qty[
n_{\rm cut}^\BSF,\,
n_{\rm growth}^\BSF (\mathtt{p})],
\label{eq:CoulombApproximationWithCutoff_Validity}
\end{align}
where we take $\mathtt{p} \sim 0.1$. In the low-velocity regime, the kinematic cutoff controls the relevant levels, with
$n_{\rm cut}^{\BSF}\to \gamma^{-1/2} = (\arho \xi/2)^{1/2} \ll \sqrt{\xi}$, i.e.~the Coulomb approximation is well controlled for perturbative $\arho$. More broadly, we confirm that  \cref{eq:CoulombApproximationWithCutoff_Validity} is satisfied in the parameter range relevant for the BSF feedback considered below, although the approximation becomes less accurate when $\arho \gtrsim 0.2$
and 
$\sqrt{2\mrho/\mu} \lesssim \vrel \lesssim \sqrt{10\arho\mrho/\mu}$,  
where $n_{\rm growth}^\BSF$ becomes large while remaining below $n_{\rm cut}^\BSF \to \infty$. 

\clearpage
\section{Elastic scattering via bound-state formation}
\label{Sec:Feedback}

\begin{figure}[b!] 
\centering \begin{tikzpicture}[
    line width=0.8pt,
    xscale=1.15,
    yscale=1.0,
    vertex/.style={circle,fill=black,inner sep=2.2pt},
    cut/.style={red,dashed,line width=1pt}
]

\node[vertex] (L) at (-1.5,0) {};
\node[vertex] (R) at ( 1.5,0) {};

\draw[fermion] (-3.2, 1.0) -- (L);
\draw[fermion] (-3.2,-1.0) -- (L);
\node[left] at (-3.2, 1.0) {$X$};
\node[left] at (-3.2,-1.0) {$X$};

\draw[fermion] (R) -- (3.2, 1.0);
\draw[fermion] (R) -- (3.2,-1.0);
\node[right] at (3.2, 1.0) {$X$};
\node[right] at (3.2,-1.0) {$X$};

\draw[double,double distance=1.2pt,line width=0.45pt]
    (L) to[out=35,in=145] (R);
\node[above] at (0,0.95)
    {${\cal B}_{n\ell}(XX^\dagger)$};

\draw[scalar]
    (L) to[out=-35,in=-145] (R);
\node[below] at (0.15,-0.72) {$\varrho$};

\draw[cut] (-0.1,-1) -- (-0.1,1);

\end{tikzpicture}
\caption{Absorptive contribution to the $XX$ self-energy kernel from BSF, $XX\to{\cal B}(XX^\dagger)+\varrho$. The double line denotes an $XX^\dagger$ bound state, and the arrows on the field lines denote the $U(1)_D$ charge flow. The on-shell cut is denoted by the red dashed line.} \label{fig:BSF_Kernel_BSFsquared} 
\end{figure}
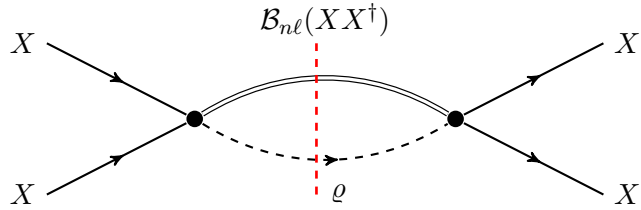

The BSF processes considered in \cref{Sec:BSFbaseline} generate an absorptive contribution to the $XX$ self-energy kernel, as illustrated in \cref{fig:BSF_Kernel_BSFsquared}. Resumming it restores consistency with the coupled unitarity constraints of \cref{eq:UnitarityConstraint_xy} and modifies both BSF and elastic scattering~\cite{Flores:2024sfy,Flores:2025uoh}. We refer to quantities obtained after this resummation as regulated, in contrast to the baseline (unregulated) quantities of \cref{Sec:BSFbaseline}. The effect of this resummation on the BSF cross sections was studied in Ref.~\cite{Flores:2026yay}. Here, we focus on its feedback on elastic scattering. This effect is particularly transparent in the model considered here: the incoming $XX$ state has no baseline long-range interaction, and we neglect the contact-like elastic contributions discussed in \cref{Sec:BSFbaseline_Model}, so its baseline elastic amplitude vanishes. Nevertheless, resumming the BSF-induced kernel generates a nonzero elastic $XX$ amplitude. In \cref{Sec:SIDM}, we use the resulting elastic scattering cross sections to study the implications for SIDM.

We focus on the feedback from the absorptive contributions to the optical potential generated by kinematically open BSF channels. The same inelastic vertices can in general also induce Hermitian dispersive contributions to the $XX$ optical potential through off-shell intermediate states~\cite{Flores:2025uoh}. Unlike the absorptive contributions, these contributions need not vanish once a BSF channel becomes kinematically closed. They may provide additional contributions to elastic scattering and further modify the BSF amplitudes. We do not include these direct dispersive contributions in the present study, leaving their investigation to future work. These direct Hermitian optical-kernel contributions should be distinguished from the principal-value dispersive contributions generated upon resummation of the absorptive kernel, encoded below in the $\WW_\ell$ matrix, which we retain.

\subsection{Unitarization}
\label{Sec:Feedback_Unitarization}

The general unitarization formalism was developed in Refs.~\cite{Flores:2024sfy,Flores:2025uoh,Flores:2026yay} and reviewed in our companion study~\cite{Petraki:2026SIDMann}, where unitarization expressions for the transfer and viscosity cross sections employed in SIDM phenomenology were also derived. We specialize from the outset to the present case, for which the baseline $XX$ phase shifts vanish, hence
$x_{\ell,\unreg}^{\elas}
=x_{\ell,\unreg}^{\T}
=x_{\ell,\unreg}^{\V}=0$.

It is convenient to define the momentum-rescaled on-shell unregulated BSF amplitudes and collect them into a vector $M_\ell^{\BSF}(k)$, with components
\begin{align}
\qty[M_\ell^{\BSF}(k)]_n
&\equiv
\qty[
\dfrac{
4k\,\omega_n(k)\Phi_n(k)
}{
\sym_\ell^{\XX}m_T^2
}
]^{1/2}
{\cal M}_{n\ell,\unreg}^{\BSF}(k)
=
\sqrt{\arho}
\qty[
\Phi_n(k)
\dfrac{1+\zeta_n^2}{\zeta_n^3}
]^{1/2}
r_{n\ell}^{\cal M}(\zeta), 
\label{eq:BSF_Mvector}
\end{align}
with $n$ running over the kinematically open bound levels, $1+\ell\leqslant n<n_{\rm cut}^{\BSF}(k)$. The matrices below act on this same set of levels. With this normalization,
\begin{align}
y_{\ell,\unreg}^{\BSF}
=
\qty(M_\ell^{\BSF})^\dagger
M_\ell^{\BSF},
\label{eq:BSF_yUnreg_Mvector}
\end{align}
consistent with \cref{eq:BSF_yUnreg}.

The momentum dependence of the BSF amplitudes renders their resummation ultraviolet convergent, so no renormalization is required~\cite{Flores:2025uoh,Flores:2026yay}, in contrast to the contact-like annihilation processes considered in the companion paper~\cite{Petraki:2026SIDMann}. Their nontrivial analytic structure in the complex momentum plane nevertheless generates finite dispersive contributions that affect the resummation. These contributions are encoded in the Hermitian matrix $\WW_\ell$, obtained from the principal-value part of the Green's function of the baseline problem projected onto the irreducible BSF amplitudes. Specializing the general result of Ref.~\cite{Flores:2026yay} to the present model gives\footnote{
Reference~\cite{Flores:2025uoh} employed a complex separable potential, which is appropriate for describing both absorptive and dispersive contributions to the 2PI kernel from contact-type interactions. The unitarization prescription was formulated by factoring the complex couplings out of the $\WW_\ell$ matrix. This prescription is also adopted in the companion paper~\cite{Petraki:2026SIDMann}. Here, we include only absorptive contributions in the separable 2PI kernel and instead absorb the couplings, without the $\im$ factor, into the $\WW_\ell$ matrix, following Ref.~\cite{Flores:2026yay}.}
%
\begin{align}
\WW_\ell^{nn'}(k)
&=
\dfrac{2\arho}{\pi}
(nn')^{\frac{3}{2}}
\qty[
\Phi_n(k)\Phi_{n'}(k)
\qty(1+\zeta_n^2)
\qty(1+\zeta_{n'}^2)
]^{\frac{1}{2}}
\, 
\PV\int_0^{+\infty}\dd z\,
\dfrac{
r_{n\ell}^{\cal M}(z)\,
r_{n'\ell}^{\cal M*}(z)
}{
z^2\qty(z^2-\zeta^2)
}.
\label{eq:BSF_Wmatrix}
\end{align}
Here, we have taken into account that the baseline $XX$ Hamiltonian does not support bound states, so only the continuum part of the spectral decomposition of its Green's function contributes. Although evaluated along the positive real axis, the principal-value integral in \cref{eq:BSF_Wmatrix} encodes the analytic structure of the BSF amplitudes in the complex $z$ plane: after analytic continuation, poles and branch cuts of the off-shell amplitudes contribute to $\WW_\ell$ through their residues and discontinuities~\cite{Flores:2025uoh,Flores:2026yay}. In practice, we evaluate the $\WW_\ell$ elements numerically along the real axis.

For the purely absorptive separable kernel considered here, the two real quantities that enter the regulated cross sections are~\cite{Flores:2025uoh,Flores:2026yay}
\begin{subequations}
\label{eq:BSF_w+wtilde}
\label[pluralequation]{eqs:BSF_w+wtilde}
\begin{align}
w_\ell^{\BSF}
&=
\qty(M_\ell^{\BSF})^\dagger
\qty(\ID+\WW_\ell^2)^{-1}
M_\ell^{\BSF},
\label{eq:BSF_w}
\\
\tilde w_\ell^{\BSF}
&=
\qty(M_\ell^{\BSF})^\dagger
\WW_\ell
\qty(\ID+\WW_\ell^2)^{-1}
M_\ell^{\BSF}.
\label{eq:BSF_wtilde}
\end{align}
\end{subequations}
Since
$x_{\ell,\unreg}^{\elas}
=x_{\ell,\unreg}^{\T}
=x_{\ell,\unreg}^{\V}=0$,
the regulated elastic, viscosity, and inclusive BSF cross sections reduce to~\cite{Petraki:2026SIDMann}
\begin{subequations}
\label{eq:RegulatedCrossSections}
\label[pluralequation]{eqs:RegulatedCrossSections}
\begin{align}
x_{\ell,\reg}^{\elas}
&=
\dfrac{
\qty(w_\ell^{\BSF})^2
+\qty(\tilde w_\ell^{\BSF})^2
}{
\qty(1+w_\ell^{\BSF})^2
+\qty(\tilde w_\ell^{\BSF})^2
},
\label{eq:RegulatedCrossSections_x}
\\
x_{\ell,\reg}^{\V}
&=
\dfrac{
\qty(w_\ell^{\BSF}-w_{\ell+2}^{\BSF})^2
+
\qty(\tilde w_\ell^{\BSF}-\tilde w_{\ell+2}^{\BSF})^2
}{
\qty[
\qty(1+w_\ell^{\BSF})^2
+\qty(\tilde w_\ell^{\BSF})^2
]
\qty[
\qty(1+w_{\ell+2}^{\BSF})^2
+\qty(\tilde w_{\ell+2}^{\BSF})^2
]
},
\label{eq:RegulatedCrossSections_xV}
\\
y_{\ell,\reg}^{\BSF}
&=
\dfrac{
w_\ell^{\BSF}
}{
\qty(1+w_\ell^{\BSF})^2
+\qty(\tilde w_\ell^{\BSF})^2
}.
\label{eq:RegulatedCrossSections_y}
\end{align}
\end{subequations}
As mentioned, for identical particles, $\sigma^{\T}=\sigma^{\elas}$, so no separate expression is required. Thus, BSF feedback generates nonzero elastic scattering despite the absence of a baseline elastic interaction.

If the $\WW_\ell$ corrections associated with the nontrivial analytic structure of the BSF amplitudes are neglected, formally taking $\WW_\ell\to0$, then
\begin{align}
w_\ell^{\BSF}
&\to
y_{\ell,\unreg}^{\BSF},
\qquad
\tilde w_\ell^{\BSF}
\to
0.
\label{eq:BSF_w_NoSingularities}
\end{align}
Note that $\tilde w_\ell^{\BSF}\to0$ in this limit because we restrict the separable kernel to its absorptive part; a Hermitian contribution to the kernel would in general yield a non-vanishing $\tilde w_\ell^{\BSF}$. 

\subsection{The effect of dispersive contributions}

The quadratic-form representation in \cref{eqs:BSF_w+wtilde} allows us to establish useful general bounds on the effect of the dispersive contributions encoded in $\WW_\ell$. Defining
\begin{align}
u_\ell
\equiv
\qty(\ID+\WW_\ell^2)^{-1/2}
M_\ell^{\BSF},
\end{align}
and using \cref{eq:BSF_yUnreg_Mvector}, we can write
\begin{align}
y_{\ell,\unreg}^{\BSF}
&=
u_\ell^\dagger
\qty(\ID+\WW_\ell^2)
u_\ell,
\qquad
w_\ell^{\BSF}
=
u_\ell^\dagger u_\ell,
\qquad
\tilde w_\ell^{\BSF}
=
u_\ell^\dagger\WW_\ell u_\ell.
\end{align}
Thus, 
$y_{\ell,\unreg}^{\BSF} = w_\ell^{\BSF} +
\lVert\WW_\ell u_\ell\rVert^2$. 
The Cauchy--Schwarz inequality then implies
\begin{align}
|\tilde w_\ell^{\BSF}|^2
&=
|u_\ell^\dagger\WW_\ell u_\ell|^2
\leqslant
\qty(u_\ell^\dagger u_\ell)
\qty(u_\ell^\dagger\WW_\ell^2u_\ell)
=
w_\ell^{\BSF}
\qty(
y_{\ell,\unreg}^{\BSF}
-
w_\ell^{\BSF}
).
\end{align}
It follows that
\begin{align}
\qty(w_\ell^{\BSF})^2
+
\qty(\tilde w_\ell^{\BSF})^2
&\leqslant
w_\ell^{\BSF}
y_{\ell,\unreg}^{\BSF},
\qquad
0
\leqslant
w_\ell^{\BSF}
\leqslant
y_{\ell,\unreg}^{\BSF}.
\label{eq:BSF_CauchySchwarz}
\end{align}
The absence of a baseline $XX$ elastic amplitude then allows us to establish a stronger relation between the full result and the $\WW_\ell=0$ approximation. Using \cref{eq:BSF_CauchySchwarz} in \cref{eq:RegulatedCrossSections_x} gives
\begin{align}
x_{\ell,\reg}^{\elas}
&\leqslant
\dfrac{
w_\ell^{\BSF}
y_{\ell,\unreg}^{\BSF}
}{
1+2w_\ell^{\BSF}
+w_\ell^{\BSF}y_{\ell,\unreg}^{\BSF}
}
\leqslant
\dfrac{
\qty(y_{\ell,\unreg}^{\BSF})^2
}{
\qty(1+y_{\ell,\unreg}^{\BSF})^2
}
=
\qty[
x_{\ell,\reg}^{\elas}
]_{\WW_\ell=0}.
\label{eq:BSF_ElasticFullVsW=0}
\end{align}
Thus, in the present model, the dispersive contributions encoded in $\WW_\ell$ can only reduce or leave unchanged the generated elastic cross section in each partial wave, relative to the $\WW_\ell=0$ approximation. This inequality relies on the vanishing baseline elastic amplitude and the purely absorptive separable kernel considered here; it does not generally extend to a nonzero baseline amplitude or additional Hermitian kernel contributions. Even under the present assumptions, no analogous inequality holds for the viscosity cross section, which depends on interference between neighboring partial waves. These distinctions will be important in our discussion of BSF-induced scattering in \cref{Sec:Feedback_VelocityDepend}.

\subsection{BSF-induced elastic scattering and its velocity dependence}
\label{Sec:Feedback_VelocityDepend}

\subsubsection{Massless emitted boson, $\WW_\ell=0$ approximation}
\label{Sec:Feedback_VelocityDepend_MasslessW=0}

\begin{figure}[t!]
\centering
\includegraphics[width=0.95\linewidth]{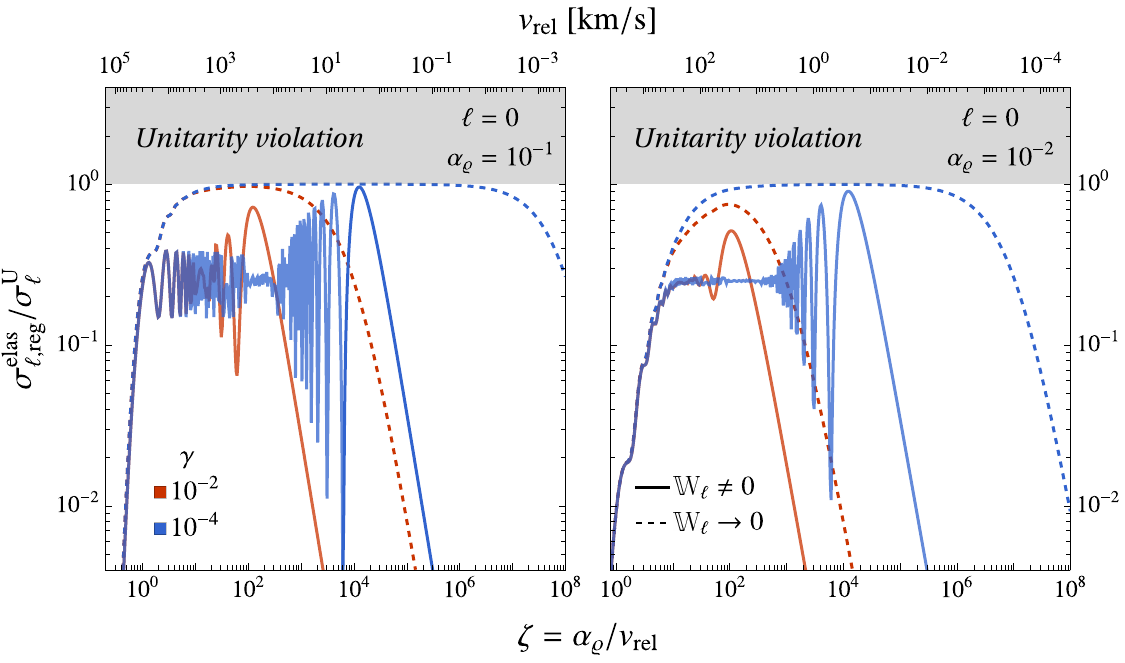}
\caption{Generated elastic cross section for $\ell=0$, as a function of $\zeta$, for representative values of $\arho$ and $\gamma=2\mrho/(\mu\arho^2)$. The full result, including the principal-value contributions encoded in $\WW_\ell$, is compared with the $\WW_\ell=0$ approximation. As established in \cref{eq:BSF_ElasticFullVsW=0}, the full result cannot exceed the $\WW_\ell=0$ result for the vanishing baseline elastic amplitude considered here.
}
\label{fig:BSF_DispersiveComparison}
\end{figure}

To build intuition for the BSF feedback on elastic scattering, we first consider the $\WW_\ell\to0$ limit, for which $w_\ell^{\BSF}\to y_{\ell,\unreg}^{\BSF}$ and $\tilde{w}_\ell^{\BSF}\to0$ according to \cref{eq:BSF_w_NoSingularities}. As long as the unregulated BSF cross section is far below its unitarity limit, the elastic scattering generated by BSF feedback is quadratically suppressed,
\begin{align}
\left.x_{\ell,\reg}^{\elas}\right|_{\WW_\ell=0}
&\simeq
\qty(y_{\ell,\unreg}^{\BSF})^2,
\qquad
\left.y_{\ell,\reg}^{\BSF}\right|_{\WW_\ell=0}
\simeq y_{\ell,\unreg}^{\BSF},
\qquad \text{for} \quad
y_{\ell,\unreg}^{\BSF} \ll 1.
\label{eq:BSFinduced_MasslessW=0_PW_smallY}
\end{align}
The feedback becomes significant once $y_{\ell,\unreg}^{\BSF}$ approaches unity. At $y_{\ell,\unreg}^{\BSF}=1$, the regulated inelastic cross section reaches its global maximum, $y_{\ell,\reg}^{\BSF}=1/4$, and the elastic cross section is correspondingly fixed to $x_{\ell,\reg}^{\elas}=1/4$. For even larger unregulated BSF cross sections, 
\begin{align}
\left.x_{\ell,\reg}^{\elas}\right|_{\WW_\ell=0} 
\simeq 1,
\qquad
\left.y_{\ell,\reg}^{\BSF}\right|_{\WW_\ell=0}
\simeq 1/y_{\ell,\unreg}^{\BSF},
\qquad \text{for} \quad
y_{\ell,\unreg}^{\BSF} \gg 1.
\label{eq:BSFinduced_MasslessW=0_PW_LargeY}
\end{align}
The correlation between the unregulated BSF rate and the generated elastic scattering can be seen by comparing \cref{fig:BSF_UniViolationAndRestoration,fig:BSF_DispersiveComparison} for small $\mrho$.

\medskip

Although $m_\varrho=0$ removes the kinematic cutoff on the partial waves, the partial-wave sums remain convergent at fixed $\zeta$. Their behavior differs substantially between the large- and low-velocity regimes, which we examine in turn.

\subsubsection*{Large velocities, $\arho\ll\vrel\ll1$}

At $\zeta\ll1$, the unregulated BSF amplitudes are suppressed, and capture is dominated by the lowest partial waves. Expanding the exact Coulombic expression in \cref{eq:CrossSectionBSF_rnl}, we find, for each fixed bound level,
\begin{align}
r_{n\ell}(\zeta)\propto\zeta^{2\ell+5},
\qquad
r_{n0}(\zeta)\simeq\frac{64\zeta^5}{n^3}.
\end{align}
Thus, summing over the bound levels, the $s$-wave contribution is
\begin{align}
y_{0,\unreg}^{\BSF}
\simeq
64\zeta_{\rm R}(3)\arho\zeta^5\ll1,
\end{align}
where $\zeta_{\rm R}$ denotes the Riemann zeta function with $\zeta_{\rm R}(3)\simeq 1.2$. Higher partial waves are suppressed by additional powers of $\zeta \ll 1$, so the $s$ wave dominates both the total elastic and viscosity cross sections. Using \cref{eq:BSFinduced_MasslessW=0_PW_smallY}, we obtain
\begin{align}
\left.\sigma_\reg^\elas\right|_{\WW_\ell=0}
&\simeq
\frac{2^{15}\pi[\zeta_{\rm R}(3)]^2\arho^{12}}
{\mu^2\vrel^{12}},
\quad
\left.\sigma_\reg^\V\right|_{\WW_\ell=0}
\simeq
\left.\frac{2\sigma_\reg^\elas}{3}\right|_{\WW_\ell=0},
\qquad
\arho\ll\vrel\ll1.
\label{eq:BSFinduced_MasslessW=0_LargeVel}
\end{align}

\subsubsection*{Low velocities, $\vrel \ll \arho$}

To estimate the angular-momentum range contributing significantly to the partial-wave sums, we consider \cref{eq:App_BSF_rEllIntegral_z} for $\gamma=0$.  At $\zeta\gg1$, the turning-point behavior of the spherical Bessel functions implies that capture is significant for $1+\ell\lesssim2\zeta$, while contributions from partial waves sufficiently above this range are exponentially suppressed.\footnote{
The piecewise Bessel approximation of Ref.~\cite{Flores:2026yay} instead suggests $1+\ell\lesssim e\zeta$. However, that approximation is not uniform for $\ell\sim\zeta$; the Bessel turning point gives the more accurate cutoff $1+\ell\lesssim2\zeta$.
}
Considering these modes, we discern two regimes: 
\begin{itemize}
\item 
For $4\arho^2 \ll \vrel \lesssim \arho$ (i.e. $\zeta \gtrsim 1$ and  $4\arho \zeta \ll 1$),  \cref{eq:App_BSF_rEllB_Massless,eq:App_BSF_rEllC_Massless} give $y_{\ell,\unreg}^\BSF \sim 4\arho\zeta \ll 1$ , thus  \cref{eq:BSFinduced_MasslessW=0_PW_smallY} applies, yielding
\begin{subequations}
\label{eq:BSFinduced_MasslessW=0_LowVel}
\label[pluralequation]{eqs:BSFinduced_MasslessW=0_LowVel}
\begin{align}
\left.\sigma_\reg^\elas\right|_{\WW_\ell=0}
&\sim
\dfrac{4\pi}{k^2} 
(4\arho\zeta)^2
\sum_{\ell\lesssim 2\zeta}
\sym_\ell^{\XX}(2\ell+1)
\approx
\dfrac{4\pi}{k^2} 
(4\arho\zeta)^2
\int_0^{2\zeta}\dd\ell\,(2\ell+1)
\nonumber \\
&\simeq
\dfrac{2^8\pi\arho^6}{\mu^2\vrel^6},
\qquad
4\arho^2 \ll \vrel \lesssim \arho. 
\label{eq:BSFinduced_MasslessW=0_Elas_LowVel_a}
\end{align}

\item
For $\vrel \lesssim 4\arho^2$, the unregulated BSF cross sections satisfy $y_{\ell,\unreg}^\BSF\gg1$ (cf.~\cref{eq:App_BSF_rEllC_Massless}), except near the upper edge of the $\ell$ range. \Cref{eq:BSFinduced_MasslessW=0_PW_LargeY} thus applies, and summing over the saturated partial waves gives
\begin{align}
\left.\sigma_\reg^\elas\right|_{\WW_\ell=0}
&\simeq
\dfrac{4\pi}{k^2}
\sum_{\ell\lesssim2\zeta}
\sym_\ell^{\XX}(2\ell+1)
\approx
\dfrac{4\pi}{k^2}
\int_0^{2\zeta}\dd\ell\,(2\ell+1)
\nonumber \\
&\simeq
\dfrac{16\pi\arho^2}{\mu^2\vrel^4},
\qquad
\vrel \lesssim 4\arho^2. 
\label{eq:BSFinduced_MasslessW=0_Elas_LowVel_b}
\end{align}
\end{subequations}
\end{itemize}
In both cases, the symmetry factor $\sym_\ell^{\XX}=2\delta_{\ell,\rm even}$ compensates, in the continuum approximation, for the spacing between the surviving even partial waves.

The viscosity cross section is instead controlled by the variation of the regulated elastic amplitudes between neighboring partial waves, as seen from \cref{eq:RegulatedCrossSections_xV,eq:BSF_w_NoSingularities} for $\WW_\ell\to0$,
\begin{align}
\left.
x_{\ell,\reg}^\V
\right|_{\WW_\ell=0}
=
\left(
\frac{y_{\ell,\unreg}^\BSF}{1+y_{\ell,\unreg}^\BSF}
-
\frac{y_{\ell+2,\unreg}^\BSF}{1+y_{\ell+2,\unreg}^\BSF}
\right)^2.
\label{eq:BSFinduced_MasslessW=0_Visc_PW}
\end{align}
Although $y_{\ell,\unreg}$ varies with $\ell$ throughout the range $\ell\lesssim2\zeta$, the differences above are strongly suppressed in the regime $\vrel \lesssim 4\arho^2$ where $y_{\ell,\unreg}\gg1$. The viscosity cross section is therefore dominated by the transition region near $\ell\simeq2\zeta$, where saturation ceases. While an analytic estimate is possible, we do not pursue it here. Numerically, we find an approximate power-law behavior in the entire low-velocity range, 
\begin{align}
\left.
\sigma_\reg^\V
\right|_{\WW_\ell=0}
\propto\vrel^{-n} 
\quad \text{with} \quad
8/3\lesssim n\lesssim 5,
\qquad
\vrel \lesssim \arho,
\tag{\ref{eq:BSFinduced_MasslessW=0_LowVel}c}
\label{eq:BSFinduced_MasslessW=0_Visc_LowVel}
\end{align}
with $n$ being lower for larger values of $\arho$. Thus, the viscosity cross section retains a strong velocity dependence, comparable to or milder than the $\vrel^{-4}$ scaling of the total elastic cross section in the saturated low-velocity regime. There is also a significant numerical suppression of $\sigma_\reg^\V$ with respect to $\sigma_\reg^\elas$.

\begin{figure}[p]
\centering
\includegraphics[width=0.95\linewidth]{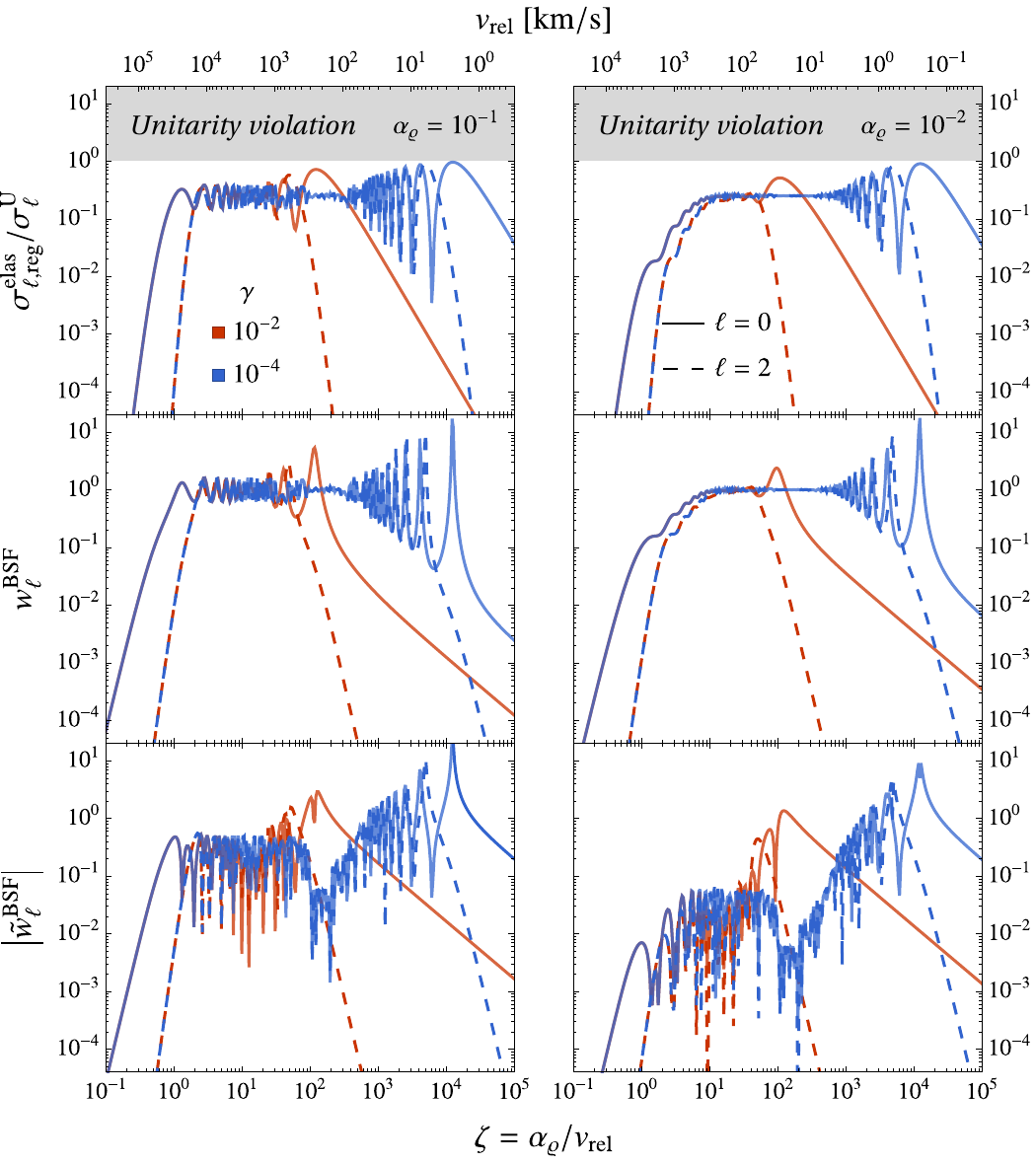}
\caption{Full BSF unitarization, including the principal-value contributions encoded in $\WW_\ell$, as a function of $\zeta$, for representative values of $\ell$, $\arho$, and $\gamma=2\mrho/(\mu\arho^2)$. The panels show the generated elastic cross section $x_{\ell,\reg}^{\elas}$ (\emph{top row}) and the regularization parameters $w_\ell^{\BSF}$ (\emph{middle row}) and $|\tilde{w}_\ell^{\BSF}|$ (\emph{bottom row}). The phase-space cutoff induced by the finite $\varrho$ mass truncates the tower of kinematically accessible bound levels entering both the absorptive and principal-value contributions and thereby modifies the low-velocity behavior.
}
\label{fig:BSF_FullUnitarization}
\end{figure}

\subsubsection{Massless emitted boson, including $\WW_\ell$}
\label{Sec:Feedback_VelocityDepend_MasslessW}

We now retain the principal-value contributions encoded in $\WW_\ell$. Their effect becomes important once the unregulated inclusive BSF cross section approaches the unitarity limit~\cite{Flores:2026yay}. The full calculation then yields $w_\ell^{\BSF}\leqslant y_{\ell,\unreg}^{\BSF}$, consistently with \cref{eq:BSF_CauchySchwarz}. The off-diagonal elements of $\WW_\ell$, which couple different intermediate bound levels, are essential for preventing $w_\ell^{\BSF}$ from following the strong low-velocity growth of $y_{\ell,\unreg}^{\BSF}$ found in the $\WW_\ell=0$ approximation~\cite{Flores:2026yay}. A comparison between the two calculations, albeit for finite $\mrho$, is shown in \cref{fig:BSF_DispersiveComparison}.

Several limits of the regulated elastic cross section in \cref{eq:RegulatedCrossSections_x}, illustrated in \cref{fig:BSF_DispersiveComparison,fig:BSF_FullUnitarization}, are useful for interpreting the resulting elastic scattering. For $y_{\ell,\unreg}^\BSF \ll 1$, $w_\ell^{\BSF}$ and $|\tilde w_\ell^{\BSF}|$ are small, and $w_\ell^{\BSF} \sim y_{\ell,\unreg}^\BSF$ while $|\tilde w_\ell^{\BSF}|/w_\ell^{\BSF}$ is suppressed by the effective radiative coupling, equal to $4\arho$ in the present model. Then, for small couplings, 
\begin{subequations}
\label{eq:BSFinduced_MasslessFull_PW}
\label[pluralequation]{eqs:BSFinduced_MasslessFull_PW}
\begin{align}
x_{\ell,\reg}^{\elas} \simeq \qty(y_{\ell,\unreg}^{\BSF})^2,
\qquad
y_{\ell,\reg}^{\BSF} \simeq y_{\ell,\unreg}^{\BSF},
\qquad
\text{for}
\quad y_{\ell,\unreg}^\BSF \ll 1, 
\quad \arho \ll 1.
\label{eq:BSFinduced_MasslessFull_PW_smallW}
\end{align}
Once $y_{\ell,\unreg}^{\BSF}\gtrsim1$, and for sufficiently small $\arho$, we find numerically that $w_\ell^{\BSF}$ tends to saturate near unity, and $\tilde{w}_\ell^{\BSF}$ remains coupling-suppressed. Consequently,

\begin{align}
x_{\ell,\reg}^{\elas}
\simeq
y_{\ell,\reg}^{\BSF}
\simeq
1/4,
\qquad
\text{for}
\quad y_{\ell,\unreg}^\BSF \gg 1, 
\quad \arho \ll 1.
\label{eq:BSFinduced_MasslessFull_PW_largeW}
\end{align}
\end{subequations}
\Cref{eq:BSFinduced_MasslessFull_PW_largeW} differs qualitatively from the $\WW_\ell=0$ approximation~\eqref{eq:BSFinduced_MasslessW=0_PW_LargeY}, which instead gives $x_{\ell,\reg}^{\elas}\to1$ and $y_{\ell,\reg}^{\BSF}\to0$ as $y_{\ell,\unreg}^{\BSF}$ grows. At sufficiently large effective radiative couplings, however, the hierarchy between $w_\ell^{\BSF}$ and $\tilde w_\ell^{\BSF}$ changes, and this limiting behavior no longer applies. For $\arho\ll1$, the principal-value contributions encoded in $\WW_\ell$ thus suppress or leave unchanged the elastic scattering in each partial wave relative to the $\WW_\ell=0$ approximation, without altering its leading velocity scaling in the limits considered above.

At large velocities, \cref{eq:BSFinduced_MasslessW=0_PW_smallY,eq:BSFinduced_MasslessW=0_LargeVel,eq:BSFinduced_MasslessFull_PW_smallW} then imply
\begin{align}
\sigma_\reg^\elas
&\sim
\dfrac{2^{15}\pi\zeta_{\rm R}(3)^2\arho^{12}}
{\mu^2\vrel^{12}},
\qquad
\sigma_\reg^\V
\sim
\dfrac{2\sigma_\reg^\elas}{3},
\qquad \text{for} \quad
\arho \ll \vrel \ll 1
.
\label{eq:BSFinduced_MasslessFull_LargeVel}    
\end{align}
At low velocities, the estimates of
\cref{eq:BSFinduced_MasslessW=0_Elas_LowVel_a,eq:BSFinduced_MasslessW=0_Elas_LowVel_b,eqs:BSFinduced_MasslessFull_PW} yield
\begin{subequations}
\label{eq:BSFinduced_MasslessFull_LowVel}
\label[pluralequation]{eqs:BSFinduced_MasslessFull_LowVel}
\begin{align}
\sigma_{\reg}^{\elas}
\sim
\begin{cases}
\dfrac{2^8\pi\arho^6}{\mu^2\vrel^6}
,~~
&4\arho^2 \ll \vrel \lesssim \arho,
\\[1em]
\dfrac{1}{4}
\left.\sigma_{\reg}^{\elas}\right|_{\WW_\ell=0}
\sim
\dfrac{4\pi\arho^2}{\mu^2\vrel^4}
,~~
&\vrel \lesssim 4\arho^2.
\end{cases}
\label{eq:BSF_MasslessFull_Elas_LowVel}
\end{align}
In this velocity range, the viscosity cross section remains sensitive to the variation of the regulated elastic amplitudes between neighboring partial waves, as seen from \cref{eq:RegulatedCrossSections_xV}. Numerically, we find that the $\WW_\ell$ contributions weaken these cancellations, allowing the viscosity cross section to exhibit a velocity dependence approaching that of its elastic counterpart,
\begin{align}
\sigma_\reg^\V 
\propto 
\vrel^{-n},
\qquad \text{with} \quad
7/2 \lesssim n \lesssim 4,
\qquad \text{for} \quad
\vrel \lesssim \arho,
\label{eq:BSFinduced_MasslessFull_Visc_LowVel}
\end{align} 
\end{subequations}
with the variation of $n$ with $\arho$ reversed with respect to the $\WW_\ell=0$ limit: here, $n$ is larger for larger $\arho$. Thus, while the $\WW_\ell$ contributions necessarily suppress or leave unchanged the total elastic cross section, as established in \cref{eq:BSF_ElasticFullVsW=0}, their effect on the viscosity cross section can be opposite. Indeed, our numerical results show a significant enhancement of $\sigma_\reg^\V$ relative to the $\WW_\ell=0$ approximation in the regime considered.

\medskip

\begin{figure}[t!]
\centering
\includegraphics[width=\linewidth]{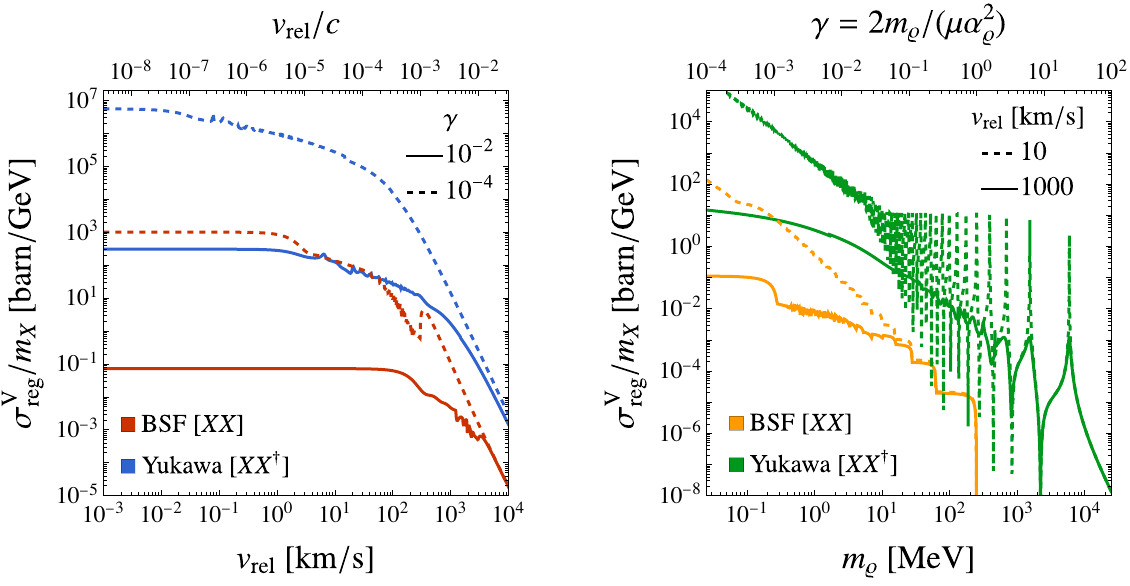}
\caption{
Total viscosity cross sections per unit DM mass, for $\arho=10^{-1}$ and $\mX=100$~GeV.
\emph{Left:} As a function of $\vrel$, for $\gamma\equiv 2\mrho/(\mu\arho^2)=10^{-4},10^{-2}$, corresponding to $\mrho=25$~keV and $2.5$~MeV, respectively.
\emph{Right:} As a function of $\gamma=2\mrho/(\mu\arho^2)$ for $\vrel=10,10^3$~km/s.
We compare BSF-induced $XX$ scattering, including the $\WW_\ell$ contributions, with Yukawa $XX^\dagger$ scattering regulated by annihilation feedback~\cite{Petraki:2026SIDMann}.
For the BSF-induced contribution, we retain the first 100 allowed even partial waves and verify convergence of the viscosity sum.
The BSF-induced contribution grows toward low velocities and smaller $\mrho$, but generally remains subleading to Yukawa scattering, except near antiresonances.
The narrowness of these antiresonances makes them difficult to resolve numerically, so only some of the expected regions of BSF dominance are visible.
}
\label{fig:SIDM_SymmetricVelocity}
\end{figure}

For sufficiently small but finite $\mrho$, the low-velocity enhancement of the $\mrho\to0$ limit persists over a finite velocity range before the phase-space cutoff on the excited-state tower curtails it, as we discuss next. This is shown in \cref{fig:SIDM_SymmetricVelocity}.

\subsubsection{Massive emitted boson}
\label{Sec:Feedback_VelocityDepend_Massive}

For finite $\mrho$, only kinematically open BSF channels enter $w_\ell^{\BSF}$, $\tilde w_\ell^{\BSF}$, and $\WW_\ell$. The phase-space cutoff progressively removes the highly excited levels responsible for the low-velocity enhancement of the absorptive and principal-value contributions, eventually curtailing the BSF-induced elastic scattering, as seen in \cref{fig:BSF_DispersiveComparison,fig:BSF_FullUnitarization}. 

\subsubsection*{Low-velocity growth}

The velocity scalings derived above for a massless emitted boson remain approximately applicable as long as the finite mediator mass does not significantly affect radiative capture. The onset of finite-mass effects can be estimated from the kinematic cutoff on the bound-level sum. For
\begin{align}
\vrel \gtrsim \sqrt{2\mrho/\mu},
\label{eq:BSF_Massive_LowVelocity_Range}
\end{align}
we have $n_{\rm cut}^\BSF=\infty$, so the tower of accessible bound levels is not truncated, although some phase-space suppression may still occur. At lower velocities, the kinematic cutoff progressively removes excited bound levels and restricts the partial waves that can contribute to BSF-induced elastic scattering. After an intermediate transition regime, the scattering becomes $s$-wave dominated and approaches a low-momentum plateau, provided that the ground-state capture channel remains open at threshold ($\gamma<1$), as discussed next.

\subsubsection*{Plateau at very low velocities}
At still lower velocities, the cutoff changes the behavior qualitatively, both in the $\WW_\ell\to0$ approximation and when the $\WW_\ell$ contributions are retained. As shown in \cref{App:BSFsums}, provided that $\gamma < (1+\ell)^{-2}$, which allows for BSF channels to remain kinematically open in the limit $k\to0$,
\begin{align}
y_{\ell,\unreg}^{\BSF}
\propto
k^{2\ell+1}
\qquad
\stackrel{\rm \cref{eq:BSF_yUnreg_Mvector}}{\Rightarrow}
\qquad
M_\ell^{\BSF}(k)
\propto
k^{\ell+1/2}.
\label{eq:BSF_Plateau_y}
\end{align}
At the same time, the matrix $\WW_\ell$ approaches a finite $k$-independent limit (cf.~\cref{eq:BSF_Wmatrix}). It then follows directly from the definitions in \cref{eqs:BSF_w+wtilde} that
\begin{align}
w_\ell^{\BSF}, \,
\tilde{w}_\ell^{\BSF}
\propto
k^{2\ell+1}.
\label{eq:BSF_Plateau_w+wtilde}
\end{align}
Since $w_\ell^{\BSF}$ and $\tilde w_\ell^{\BSF}$ become small, the regulated cross-section formulas in \cref{eq:RegulatedCrossSections_x,eq:RegulatedCrossSections_y} give
\begin{subequations}
\label{eq:BSF_LowVelocity}
\label[pluralequation]{eqs:BSF_LowVelocity}
\begin{align}
x_{\ell,\reg}^{\elas}
&\propto
k^{4\ell+2},
&
\sigma_{\ell,\reg}^{\elas}
&\propto
k^{4\ell},
\label{eq:BSF_Plateau_Elastic}
\\
y_{\ell,\reg}^{\BSF}
&\propto
k^{2\ell+1},
&
\sigma_{\ell,\reg}^{\BSF}\vrel
&\propto
k^{2\ell}.
\label{eq:BSF_Plateau_Inelastic}
\end{align}
\end{subequations}
The BSF-induced elastic scattering therefore eventually exhibits the usual low-velocity behavior of a short-range interaction: the $s$-wave cross section approaches a constant, while higher partial waves are velocity suppressed.

In the smooth approximation of \cref{App:BSFsums}, the asymptotic regime that gives rise to \cref{eqs:BSF_LowVelocity} corresponds parametrically to
$\zeta\gtrsim (\gamma b_\ell)^{-1}$ (region A).
For the $s$ wave, provided $\gamma<1$ so that the $n=1$ capture channel remains open as $k\to0$, the onset of the asymptotic plateau is estimated to occur at
\begin{align}
\vrel \lesssim
\dfrac{\mrho}{\mu\arho}.
\label{eq:BSFelasticscatt_PlateauVelocity}
\end{align}
This explains the low-velocity flattening observed in \cref{fig:SIDM_SymmetricVelocity}. The height of the $s$-wave plateau can also be estimated. In the $\vrel\to0$ limit, \cref{eq:App_BSF_LargeZetaCoefficient_sWave} implies
\begin{align}
y_{0,\unreg}^{\BSF}
\simeq
A_0(\gamma)\,\vrel,
\qquad
A_0(\gamma)
=
64\sum_{n=1}^{1/\sqrt{\gamma}}
n^3\sqrt{1-\gamma^2n^4}.
\end{align}
For $\gamma\ll1$, approximating the sum by an integral gives $A_0(\gamma)\simeq32/(3\gamma^2)$.

For fixed $\gamma$, the low-energy $\WW_0$ matrix scales with $\arho$, so that its corrections become negligible at sufficiently small coupling. In this regime, $x_{0,\reg}^\elas \sim (y_{0,\unreg}^{\BSF})^2$, thus
\begin{subequations}
\begin{align}
\sigma_{\reg}^{\elas}
\simeq
\sigma_{0,\reg}^{\elas}
~\stackrel{\WW_0\to 0}{\sim}~
\sym_0^{\XX}
\,\dfrac{4\pi}{\mu^2}
\,A_0^2(\gamma)
\simeq
\dfrac{2^{13}\pi}{3^2\mu^2\gamma^4}
=
\dfrac{2^{9}\pi}{3^2}
\dfrac{\mu^2\arho^8}{\mrho^4},
\quad
\text{very small}~\arho .
\label{eq:BSFelasticscatt_PlateauHeight_W=0}
\end{align}
As $\arho$ increases, the $\WW_0$ contributions become more important. The plateau remains velocity independent, but its height is reduced relative to the $\WW_0=0$ result, consistently with \cref{eq:BSF_ElasticFullVsW=0}. Numerically, we find
\begin{align}
\sigma_{\reg}^{\elas}
\simeq
\sigma_{0,\reg}^{\elas} \propto
\dfrac{\arho^2}{\mrho^2},
\qquad
\text{larger}~\arho.
\label{eq:BSFelasticscatt_PlateauHeight_withW}
\end{align}
\end{subequations}
Because higher partial waves are suppressed in this regime, 
$\sigma_\reg^\V = (2/3)\sigma_\reg^\elas$.

\subsubsection{Comparison with nonresonant Yukawa scattering}
\label{Sec:Feedback_VelocityDepend_Yukawa-BSF_nonreson}

\paragraph{Massless mediator.}

The massless-mediator limit highlights an important distinction between BSF-induced and Coulomb scattering. For BSF-induced scattering, capture is dynamically suppressed for $\ell \gtrsim 2\zeta$, so the partial-wave sums remain finite at fixed nonzero velocity, as seen in \cref{eq:BSFinduced_MasslessFull_LargeVel,eq:BSFinduced_MasslessFull_LowVel}.
No analogous large-$\ell$ cutoff exists in Coulomb scattering. Introducing $\ell_{\max}$ as an infrared regulator, for $\ell_{\max}\gg\zeta\gg1$ one finds parametrically 
\begin{align}
\sigma_{\rm Coul}^{\elas}
\sim
\dfrac{\ell_{\max}^2}{\mu^2\vrel^2}
\qquad \text{and} \qquad
\sigma_{\rm Coul}^{\V}
\sim
\dfrac{\arho^2}{\mu^2\vrel^4}
\ln\qty(\dfrac{\ell_{\max}}{\zeta}),
\label{eq:CoulombScattering}
\end{align}
up to an oscillatory order-one factor in the elastic cross section. While the BSF-induced partial-wave sums remain finite, the Coulomb cross sections are infrared divergent.

\paragraph{Massive mediator.} For finite $\mrho$, the BSF-induced scattering resembles Yukawa scattering more closely: in both cases the viscosity cross section can grow strongly toward low velocities before approaching a plateau. However, the parametric behavior differs.

In the non-perturbative regime, $\xi \gg 1$ or $\mu \arho \gg \mrho$, the partial waves contributing significantly to Yukawa scattering may be estimated classically as
\begin{align}
\ell^{\rm Yuk}
\sim 
\ell_{\rm clas}
\sim
\frac{k}{\mrho}
=\frac{\mu\vrel}{\mrho},
\qquad
\mu\vrel\gtrsim\mrho.
\label{eq:Yukawa_PartialWaveRange}
\end{align}
At $\vrel\ll\arho$, provided that $\ell^{\Yuk}\gg\zeta$, the Coulomb estimates of \cref{eq:CoulombScattering} apply parametrically, with $\ell_{\max}$ replaced by $\ell^{\Yuk}$. This gives
\begin{align}
\sigma_\Yuk^\elas \propto \dfrac{1}{\mrho^2}
\quad \text{and} \quad
\sigma_\Yuk^\V \propto 
\dfrac{\arho^2}{\mu^2\vrel^4}
\ln \left( \dfrac{\mu \vrel^2}{\mrho \arho} \right), 
\qquad
\sqrt{\dfrac{\arho\mrho}{\mu}} 
\ll \vrel 
\ll \arho. 
\label{eq:YukawaScattering}
\end{align}
This estimate ceases to be valid at $\ell^{\rm clas} \lesssim \zeta$, or  $\vrel\lesssim\sqrt{\arho\mrho/\mu}$, where the finite range of the Yukawa interaction increasingly affects the contributing partial waves. In the intermediate regime $\mrho/\mu\lesssim\vrel\lesssim\sqrt{\arho\mrho/\mu}$, the velocity dependence is generally nontrivial and can exhibit oscillatory features associated with the Yukawa phase shifts. As $\vrel$ decreases below $\mrho/\mu$, higher partial waves become suppressed and, away from resonances, the cross sections approach an $s$-wave plateau. Estimating its height from the Coulomb $s$-wave cross section at the crossover scale $k\sim\mrho$ gives, up to the oscillatory Coulomb phase-shift factor,
\begin{align}
\sigma_{0}^{\elas}
\simeq
\dfrac{3}{2} \sigma_{0}^{\V}
\sim 
[\sigma_0^\U (k = \mrho)]_{XX^\dagger}
= \dfrac{4\pi}{\mrho^2},
\qquad
\vrel \lesssim \dfrac{\mrho}{\mu}.
\label{eq:YukawaElasticScatt_PlateauHeight}
\end{align}

\medskip

We summarize the comparison in \cref{tab:Yukawa-BSF_comparison}. Both Yukawa and BSF-induced scattering can exhibit strong enhancements toward low velocities before approaching a plateau at sufficiently low velocities. Their velocity dependence, however, differs: BSF-induced scattering is more strongly suppressed at large velocities, while its long-range-like enhancement is restricted to a parametrically narrower velocity interval. Moreover, its low-velocity plateau is suppressed by higher powers of $\arho$ at fixed masses. Consequently, BSF-induced scattering is typically subdominant to Yukawa scattering.

We reiterate that these estimates assume a vanishing baseline $XX$ potential and a perturbative radiative coupling for BSF-induced scattering, and apply away from Sommerfeld resonances and cross-section zeros for Yukawa scattering. Relaxing these assumptions can alter the relative importance of the two contributions. Of particular interest are Yukawa antiresonances, to which we now turn.

\newcolumntype{L}{>{\raggedright\arraybackslash}m{0.27\textwidth}}
\newcolumntype{M}{>{\centering\arraybackslash}m{0.30\textwidth}}
\begin{table}[htb]
\centering
\renewcommand{\arraystretch}{1.5}
\setlength{\extrarowheight}{5pt}
\begin{tabular}{L|M|M}
\hline\hline
&
\shortstack{
BSF-induced $XX$\\[1ex]
$\mrho\ll\mu\arho^2/2$
}
&
\shortstack{
Yukawa $XX^\dagger$\\[1ex]
$\mrho\ll\mu\arho$,\\
away from resonances\\
and elastic zeros
}
\\
\hline
Velocity range
&
$\arho\ll\vrel\ll1$
&
$\arho\ll\vrel\ll1$
\\[1.6ex]
$\sigma^\V$ scaling
&
$\dfrac{\arho^{12}}{\mu^2\vrel^{12}}$
&
$\dfrac{\arho^2}{\mu^2\vrel^4}$
\\
\hline
Velocity range
&
$\sqrt{\dfrac{2\mrho}{\mu}}\ll\vrel\lesssim\arho$
&
$\sqrt{\dfrac{\arho\mrho}{\mu}}\ll\vrel\ll\arho$
\\[1.6ex]
$\sigma^\V$ scaling
&
$\vrel^{-n}$,~~~$3.5\lesssim n\lesssim4$
&
$
\dfrac{\arho^2}{\mu^2\vrel^4}
$
\\
\hline
Plateau onset, $\vrel$
&
$\dfrac{\mrho}{\mu\arho}$
&
$\dfrac{\mrho}{\mu}$
\\[1.6ex]
Plateau height, $\sigma^\V$
&
$\vcenter{\hbox{\shortstack{
$\dfrac{\mu^2\arho^8}{\mrho^4}$\\
(smaller $\arho$)
}}}
\;\mbox{to}\;
\vcenter{\hbox{\shortstack{
$\dfrac{\arho^2}{\mrho^2}$\\
(larger $\arho$)
}}}$
&
$\dfrac{1}{\mrho^2}$
\\[1.6ex]
\hline\hline
\end{tabular}
\caption{
Parametric comparison of BSF-induced and nonresonant Yukawa viscosity cross sections. The intermediate scaling for the  BSF-induced scattering is inferred numerically, including the $\WW_\ell$ contributions. Logarithmic factors for Yukawa scattering and crossover regions are not shown. The plateau scaling of BSF-induced scattering for larger $\arho$ is also inferred numerically.
}
\label{tab:Yukawa-BSF_comparison}
\end{table}

\subsubsection{Comparison near Yukawa antiresonances}
\label{Sec:Feedback_VelocityDepend_Yukawa-BSF_Antiresonances}

For an attractive Yukawa potential, the baseline $s$-wave elastic cross section exhibits velocity-dependent zeros that approach discrete values of $\xi$ in the zero-momentum limit. These antiresonances arise from the Ramsauer--Townsend effect~\cite{sakurai2020modern}, and their properties and velocity dependence are discussed in more detail in Refs.~\cite{Flores:2026num,Petraki:2026SIDMann}. Near an antiresonance, the baseline $XX^\dagger$ elastic amplitude is strongly suppressed, while higher partial waves remain negligible at sufficiently low velocities due to the centrifugal barrier. However, inelastic feedback can lift these zeros by regenerating a nonzero elastic amplitude. In the present model,  $XX^\dagger\to\varrho\varrho^\dagger$ annihilation remains nonzero at the baseline elastic-scattering zeros and thereby lifts the Yukawa antiresonances through unitarization~\cite{Petraki:2026SIDMann}.

\begin{figure}[t!]
\centering
\includegraphics[width=0.95\linewidth]{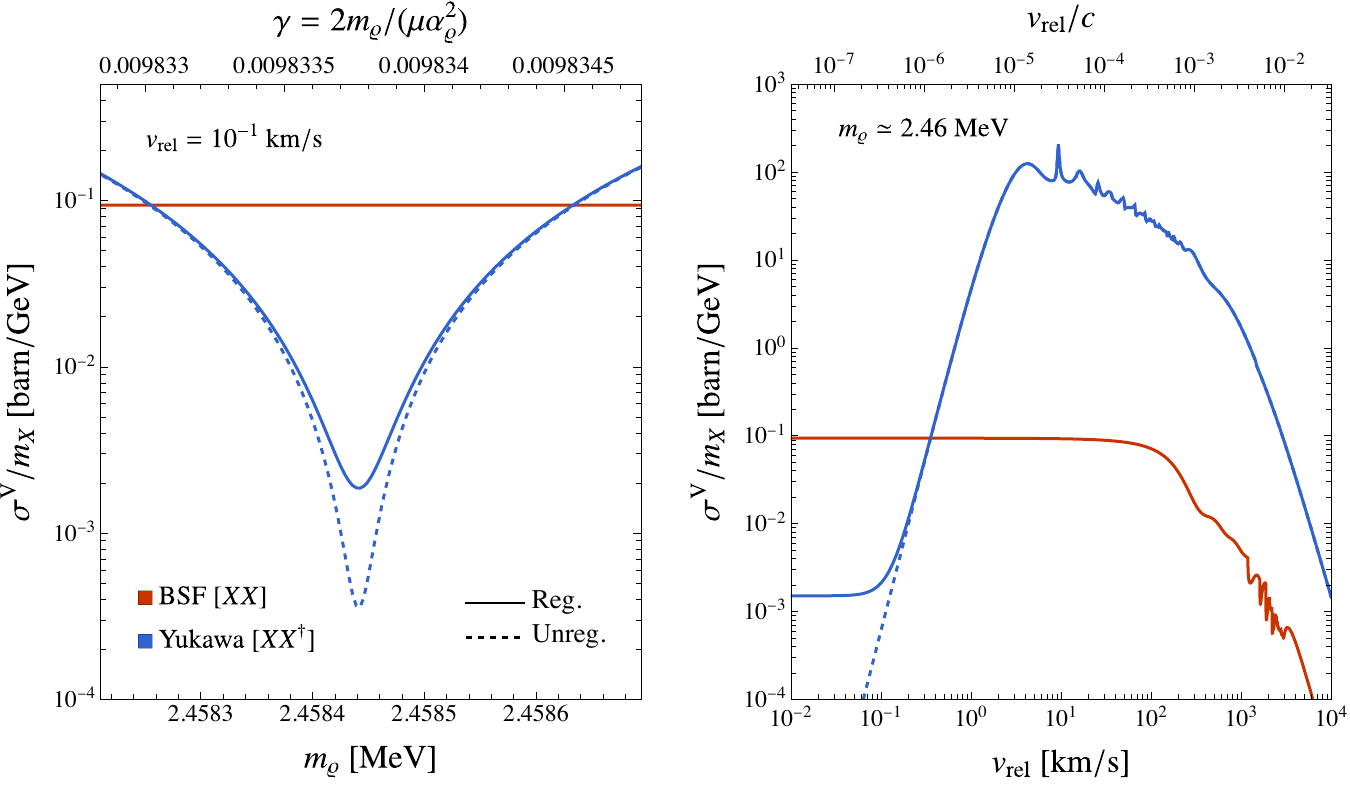}
\caption{
Viscosity cross sections per unit DM mass near an $s$-wave Yukawa antiresonance, for $\arho=10^{-1}$ and $m_X=100~{\rm GeV}$.
\emph{Left:} Cross sections as a function of $\mrho$ around the antiresonance, at $\vrel=10^{-1}~{\rm km/s}$.
\emph{Right:} Cross sections as a function of $\vrel$ at the antiresonant value of $\mrho$.
We compare the unregulated and annihilation-regulated $XX^\dagger$ Yukawa cross sections with the BSF-induced $XX$ contribution.
Near the antiresonance, the BSF-induced scattering can dominate over Yukawa scattering, despite the lifting of the latter's elastic-scattering zero by annihilation feedback, $XX^\dagger\to\varrho\varrho^\dagger$. This reverses the hierarchy between Yukawa and BSF-induced scattering found at generic parameter points.
}
\label{fig:Antiresonance_Yukawa_vs_BSF}
\end{figure}

Nevertheless, sufficiently close to an antiresonance, the BSF-induced $XX$ scattering can exceed the annihilation-regulated $XX^\dagger$ Yukawa contribution, as illustrated in \cref{fig:Antiresonance_Yukawa_vs_BSF}. The usual hierarchy between the two contributions is therefore reversed, in contrast to the behavior at generic points in parameter space, such as those shown in \cref{fig:SIDM_SymmetricVelocity}.

Despite this reversal, antiresonant regions do not typically yield phenomenologically viable SIDM. As the velocity increases away from the antiresonance, the Yukawa contribution recovers and eventually overtakes the BSF-induced one. Consequently, parameter points where BSF-induced scattering provides sizable self-interactions at subhalo or dwarf-galaxy velocities tend to exhibit even larger Yukawa cross sections at cluster velocities, in conflict with cluster-scale constraints.

\clearpage
\section{Self-interacting dark matter phenomenology}
\label{Sec:SIDM}

We now investigate the implications of the BSF-induced elastic scattering derived in \cref{Sec:Feedback} for SIDM phenomenology. We consider two different possibilities for the DM composition:
(i) symmetric DM, consisting of equal populations of $X$ and $X^\dagger$, where the BSF-induced $XX$ scattering contributes alongside the baseline Yukawa scattering of $XX^\dagger$, and
(ii) asymmetric DM~\cite{Petraki:2013wwa,Zurek:2013wia}, consisting only of $X$, such that, under the assumptions of \cref{Sec:BSFbaseline}, elastic self-scattering is generated entirely by BSF feedback. We focus here on the elastic momentum and energy transport induced by this feedback. BSF itself is an inelastic process and may additionally affect halo evolution through energy transfer and the production of bound states~\cite{Rosenlyst:2026koq}; a treatment of these effects is beyond the scope of this work.

\subsection{Halo-averaged cross sections and SIDM benchmarks}
\label{Sec:SIDM_AveragedWeightedCrossSections}

To identify regions relevant for SIDM phenomenology, we use the representative benchmarks at galaxy-cluster, dwarf-galaxy and subhalo scales, as in the companion study~\cite{Petraki:2026SIDMann},
\begin{subequations}
\label{eq:SIDM_Benchmarks}
\label[pluralequation]{eqs:SIDM_Benchmarks}
\begin{align}
\dfrac{\overline{\sigma}_{\SIDM}^{\V}}{\mX}
&\lesssim
1~{\rm barn\,GeV^{-1}},
&&
\text{at}~
v_c \sim 10^3~{\rm km/s},
&&
\text{galaxy-cluster scale},
\label{eq:SIDM_Benchmark_Cluster}
\\
\dfrac{\overline{\sigma}_{\SIDM}^{\V}}{\mX}
&\gtrsim
1~{\rm barn\,GeV^{-1}},
&&
\text{at}~
v_c \sim 10~{\rm km/s},
&&
\text{dwarf-galaxy scale},
\label{eq:SIDM_Benchmark_Dwarf}
\\
\dfrac{\overline{\sigma}_{\SIDM}^{\V}}{\mX}
&\sim
10^2~{\rm barn\,GeV^{-1}},
&&
\text{at}~
v_c \sim 1~{\rm km/s},
&&
\text{subhalo scale},
\label{eq:SIDM_Benchmark_Subhalos}
\end{align}
\end{subequations}
where the halo-averaged viscosity cross section is defined via 
\begin{align}
    \overline{\sigma}^{\V}
\equiv
\dfrac{\average{\sigma^{\V}\vrel^3}}
{4\average{v_{\DM}^2}\average{\vrel}},
\end{align}
as detailed in \cite{Colquhoun:2020adl,Petraki:2026SIDMann}. We evaluate the averages assuming an isotropic Maxwellian velocity distribution truncated at $\vesc\simeq\sqrt{2}v_c$, following the prescription derived in Ref.~\cite[Sec.~2.5 and App.~B]{Petraki:2026SIDMann}. Here, $v_c/\sqrt{2}$ is the one-dimensional velocity dispersion of the corresponding untruncated Maxwellian.

\subsection{Symmetric dark matter: BSF feedback versus Yukawa scattering}
\label{Sec:SIDM_SymmetricDM}

For symmetric DM, the halo contains equal populations of $X$ and $X^\dagger$. Assuming that the two species follow the same velocity distribution and taking into account that $\sigma_{\XX}^{\V}=\sigma_{\XdaggerXdagger}^{\V}$, the halo-averaged viscosity cross section is
\begin{align}
\overline{\sigma}_{\SIDM}^{\V}
=
\dfrac{1}{2}
\qty(
\overline{\sigma}_{\XXdagger}^{\V}
+
\overline{\sigma}_{\XX}^{\V}
).
\label{eq:SIDM_SymmetricCrossSection}
\end{align}
The two contributions have different origins. The $XX^\dagger$ elastic scattering is generated by the Yukawa interaction, including the feedback from $XX^\dagger\to\varrho\varrho^\dagger$ annihilation studied in Ref.~\cite{Petraki:2026SIDMann}. By contrast, the baseline $XX$ elastic amplitude vanishes under the assumptions of \cref{Sec:BSFbaseline}, and the corresponding scattering is generated entirely by BSF feedback, as studied in \cref{Sec:Feedback_VelocityDepend}.

\begin{figure}[p]
\centering
\includegraphics[width=0.90 \linewidth]{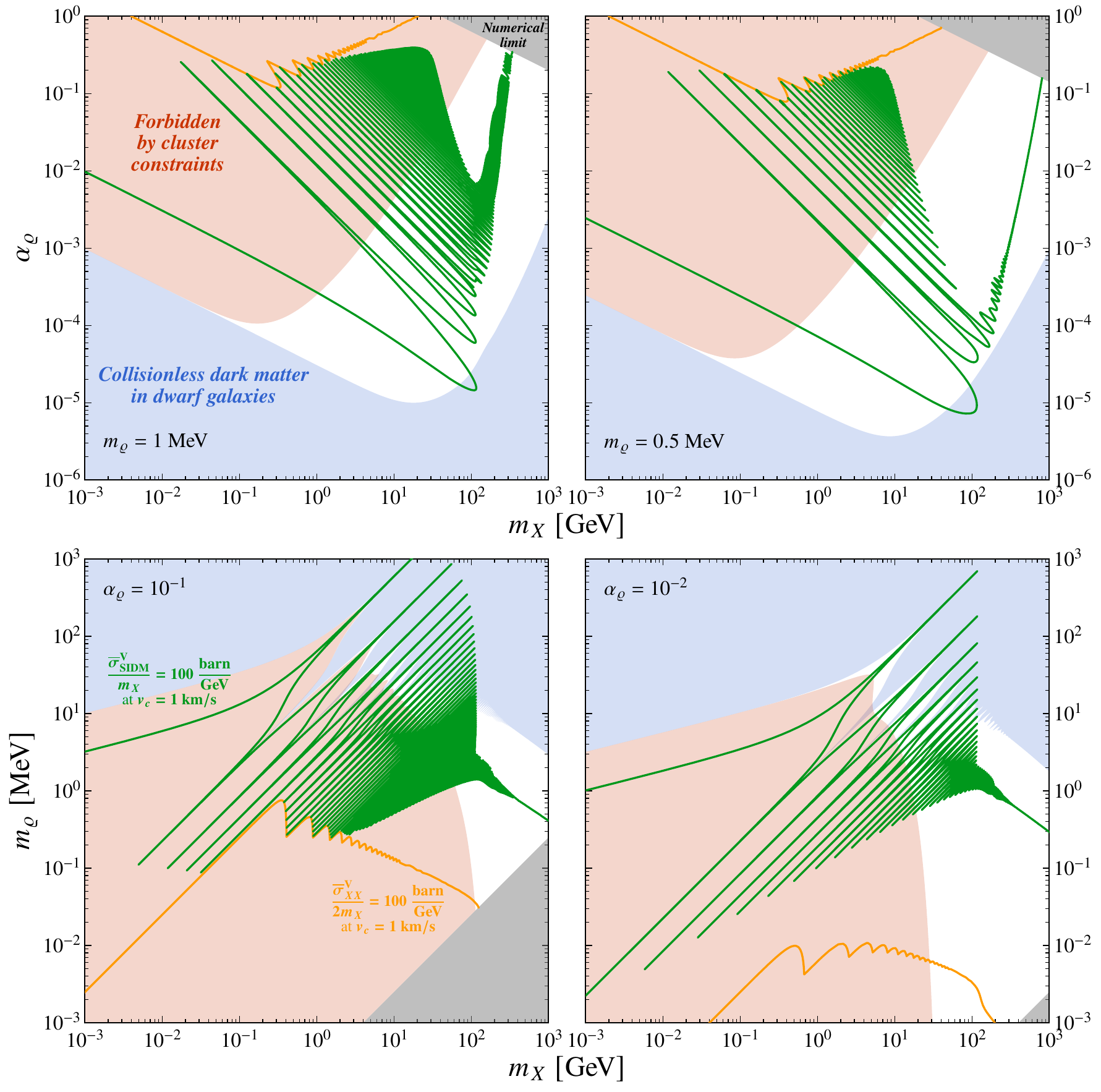}
\caption{
Halo-averaged viscosity cross section per unit DM mass, $\overline{\sigma}_{\rm SIDM}^{\V}/\mX$, for symmetric DM. We show the DM mass--coupling plane for fixed mediator masses (\emph{upper row}) and the DM mass--mediator mass plane for fixed couplings (\emph{lower row}). The red-shaded regions exceed the cluster-scale bound of \cref{eq:SIDM_Benchmark_Cluster}, while the blue-shaded regions fall below the dwarf-galaxy benchmark of \cref{eq:SIDM_Benchmark_Dwarf}. The white regions satisfy both criteria. The green contours correspond to the subhalo-scale benchmark of \cref{eq:SIDM_Benchmark_Subhalos}. These results include both Yukawa $XX^\dagger$ scattering and BSF-induced $XX$ scattering, with the relative weights appropriate for symmetric DM. The orange contours correspond to the subhalo benchmark \eqref{eq:SIDM_Benchmark_Subhalos}, but retaining only the BSF-induced scattering. Where the orange and green contours coincide, BSF feedback dominates,  curtailing the Yukawa antiresonance pattern. At other antiresonances, annihilation feedback is more important in lifting the suppression of Yukawa scattering~\cite{Petraki:2026SIDMann}. The intricate structure of the green contours at large $\arho$ and $\mX$ is due to the increasing importance of higher partial waves. The gray regions marked ``Numerical limit'' correspond to parameters for which the large number of contributing bound levels renders the full calculation numerically challenging. 
}
\label{fig:SIDM_SymmetricParameterSpace}
\end{figure}

The resulting SIDM parameter space is shown in \cref{fig:SIDM_SymmetricParameterSpace}, using the benchmarks of \cref{eqs:SIDM_Benchmarks}. The shading identifies regions exceeding the cluster-scale benchmark or exhibiting insufficient scattering at dwarf-galaxy velocities. The green contours correspond to the subhalo-scale benchmark for the full symmetric-DM cross section. As anticipated from the comparison in \cref{Sec:Feedback_VelocityDepend}, Yukawa scattering generally dominates the elastic transport of symmetric DM. Since the subhalo benchmark probes very low velocities, the green contours exhibit a pronounced resonance and antiresonance structure associated with the nonperturbative $XX^\dagger$ Yukawa interaction.

To assess the importance of BSF feedback, we also show in orange the subhalo-scale contours of \cref{eq:SIDM_Benchmark_Subhalos}, obtained by retaining only the BSF-induced $XX$ contribution in \cref{eq:SIDM_SymmetricCrossSection}. Where the orange and green contours coincide, BSF-induced scattering dominates the total, effectively filling the suppression of Yukawa scattering near antiresonances. At other antiresonances, annihilation feedback plays a more important role in lifting the baseline Yukawa elastic-scattering zeros, while the BSF-induced contribution remains subdominant. The impact of annihilation feedback on the symmetric-DM parameter space is examined separately in Ref.~\cite[Fig.~15]{Petraki:2026SIDMann}, which compares the Yukawa scattering predictions with and without this feedback.

Although BSF-induced scattering can locally dominate near Yukawa antiresonances, these regions do not generally open new parameter space satisfying the SIDM benchmarks. In particular, parameter choices yielding sufficiently large BSF-induced scattering at low velocities tend to produce excessive Yukawa scattering at cluster velocities, as discussed in \cref{Sec:Feedback_VelocityDepend_Yukawa-BSF_Antiresonances}. Thus, in the parameter range explored here, BSF feedback modifies the antiresonance structure of symmetric-DM self-scattering without substantially altering its phenomenologically viable parameter space.

\subsection{Asymmetric dark matter: BSF-generated self-interactions}
\label{Sec:SIDM_AsymmetricDM}

We now consider the fully asymmetric limit, in which the DM population in the halo consists entirely of $X$ particles. The $XX^\dagger$ Yukawa-scattering channel is then absent, and the halo-averaged viscosity cross section relevant for elastic transport is
\begin{align}
\overline{\sigma}_{\SIDM}^{\V}
=
\overline{\sigma}_{\XX}^{\V}.
\label{eq:SIDM_AsymmetricCrossSection}
\end{align}
The resulting self-interactions can be strongly velocity dependent, as discussed in \cref{Sec:Feedback_VelocityDepend}. The cross section can grow toward lower velocities until the finite mediator mass curtails this enhancement, eventually approaching an $s$-wave plateau if capture remains kinematically allowed. The velocity range over which this growth develops determines the relative strength of the interaction in clusters, dwarf galaxies, and smaller subhalos. BSF feedback can therefore generate sizable self-interactions on galactic and subgalactic scales while maintaining sufficiently weak scattering at cluster velocities.

\begin{figure}[p]
\centering
\includegraphics[width=\linewidth]{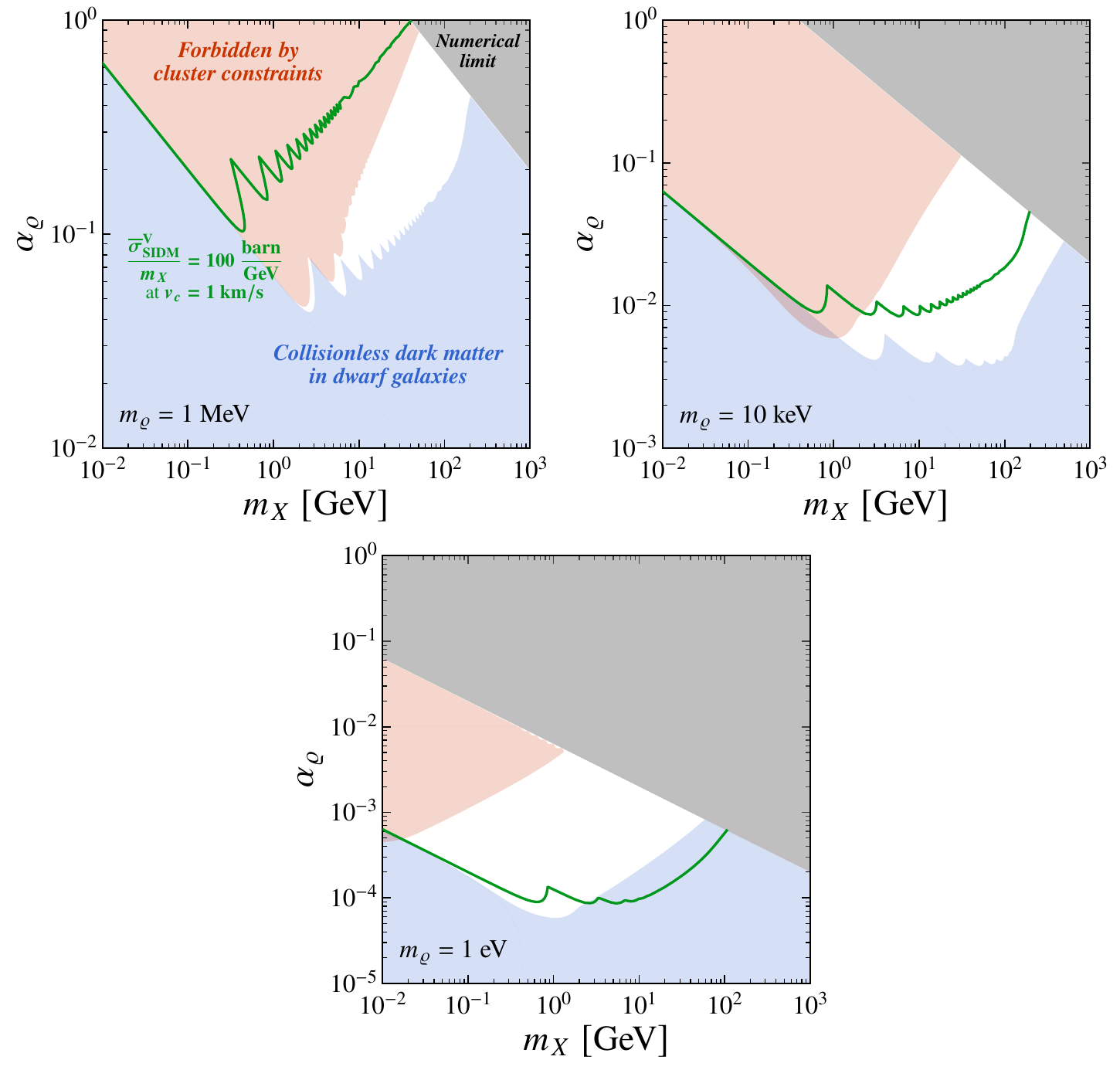}
\caption{
Halo-averaged viscosity cross section per unit DM mass, $\overline{\sigma}_{\rm SIDM}^{\V}/m_X$, for asymmetric DM, in the DM mass--coupling plane, for fixed mediator masses as indicated. The elastic scattering is generated entirely by BSF feedback, including the principal-value contributions encoded in $\WW_\ell$. Color coding as in \cref{fig:SIDM_SymmetricParameterSpace}.
}
\label{fig:SIDM_AsymmetricParameterSpace}
\end{figure}

\begin{figure}[p]
\centering
\includegraphics[width=\linewidth]{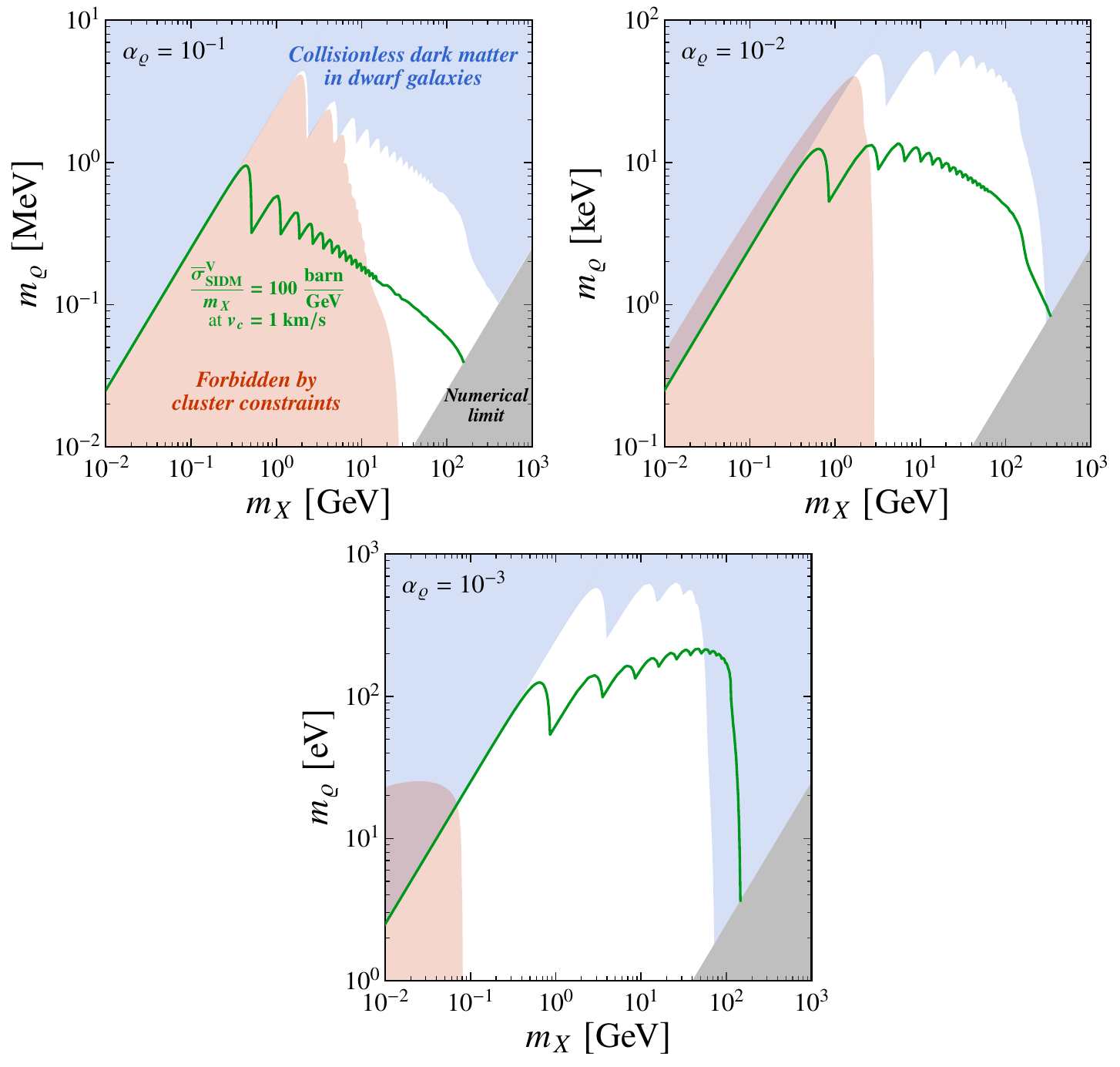}
\caption{
As in \cref{fig:SIDM_AsymmetricParameterSpace}, in the DM mass-mediator mass plane, for fixed values of the coupling $\arho$ as indicated.  
}
\label{fig:SIDM_AsymmetricParameterSpace_med_VS_mu}
\end{figure}

We illustrate these possibilities in
\cref{fig:SIDM_AsymmetricParameterSpace,fig:SIDM_AsymmetricParameterSpace_med_VS_mu},
using the cluster-, dwarf-, and subhalo-scale benchmarks of \cref{eqs:SIDM_Benchmarks}. The panels exhibit three qualitatively distinct regimes. In the first, scattering can be sizable at dwarf-galaxy velocities while satisfying the cluster constraint, but falls short of the larger subhalo-scale cross section within this parameter space. In the second, all three benchmarks can be met: the scattering is weak enough at cluster scales, sizable at dwarf scales, and strongly enhanced at subhalo scales. In the third, the enhancement becomes significant only below dwarf-galaxy velocities, allowing strong subhalo-scale scattering to coexist with essentially collisionless DM in dwarf galaxies. Thus, BSF feedback can produce qualitatively distinct self-interaction patterns across halo scales. The contours include the full unitarized result, with principal-value contributions in $\WW_\ell$. The ``Numerical limit'' marks computationally challenging regions due to numerous bound levels, not physical exclusions.


\subsection{Cosmological and astrophysical considerations}
\label{Sec:SIDM_Cosmology}

We briefly comment on several cosmological and astrophysical considerations relevant to the scenarios discussed above. Their detailed implications depend on the thermal history of the dark sector and on interactions beyond those required for the scattering dynamics studied here, and are therefore beyond the scope of this work.

\subsubsection{Dark-matter relic density}
\label{Sec:SIDM_Cosmology_RelicDensity}

Assuming a standard thermal history in which the dark sector was populated and internally thermalized, the DM abundance is set by thermal freeze-out and the potential presence of an asymmetry. In the present model, depletion can receive important contributions from Sommerfeld-enhanced annihilation and from the exceptionally efficient BSF processes mediated by charged-scalar emission~\cite{Oncala:2019yvj,Petraki:2025zvv}. Including these effects, Ref.~\cite{Petraki:2025zvv} finds $\arho^{\rm sym}\simeq3\times10^{-6}$ at $\mX\simeq1~{\rm GeV}$, assuming a common dark- and visible-sector temperature at DM decoupling. 
We adopt the numerical results of Ref.~\cite{Petraki:2025zvv} for the masses studied there and extrapolate to lower masses using the local power law scaling, $\arho^{\rm sym}\propto\mX^{0.54}$, inferred from these results around $\mX \sim$~GeV.

\begin{figure}[t!]
\centering
\includegraphics[width=0.85\linewidth]{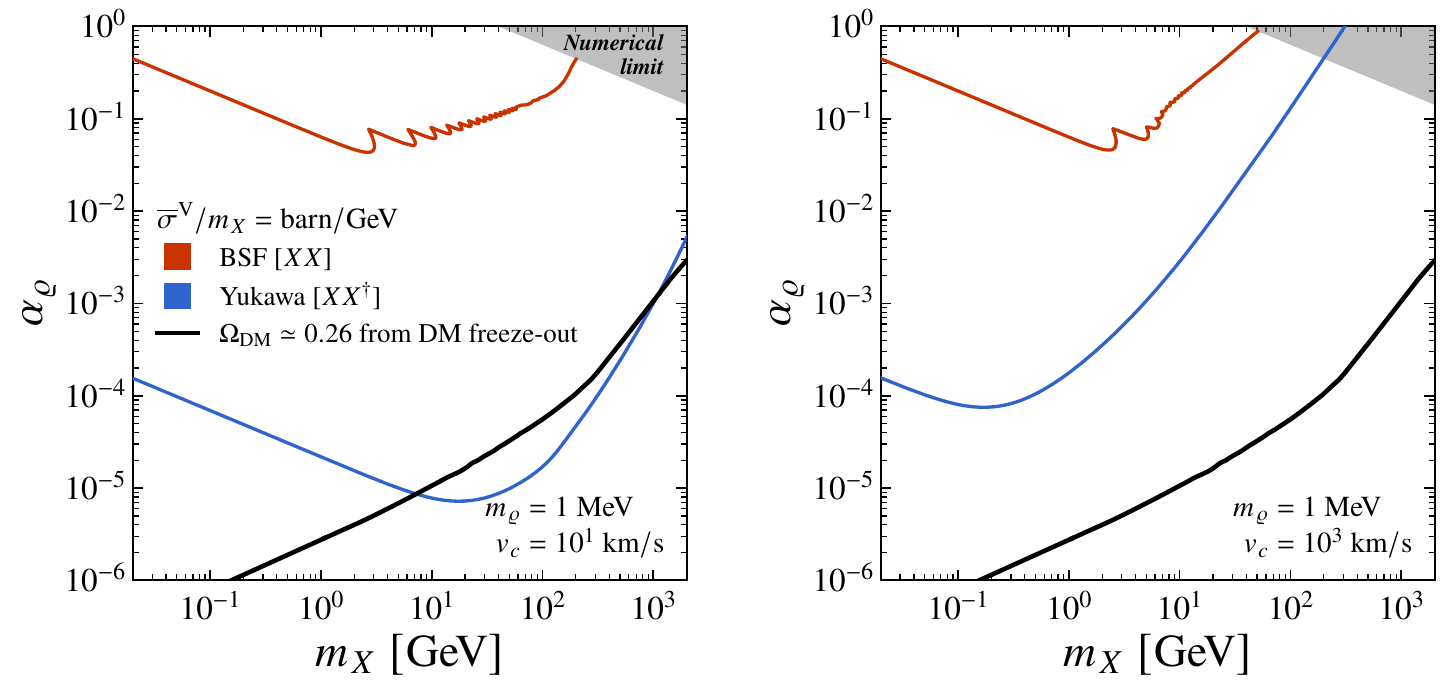}
\caption{
\emph{Left:} Working estimate $\arho^{\rm sym}(\mX)$ for the coupling reproducing the observed symmetric thermal-relic abundance~\cite{Petraki:2025zvv}; larger couplings may accommodate asymmetric thermal relics. Colored contours show $\overline{\sigma}^{\V}/\mX=1~{\rm barn/GeV}$ for $XX^\dagger$ Yukawa scattering and for BSF-induced elastic $XX$ scattering, at $v_c=10~{\rm km/s}$.
\emph{Right:} Same quantities at $v_c=10^3~{\rm km/s}$.
In both panels, the cross section generally grows towards larger couplings and smaller masses, though resonances may introduce non-monotonicity.
}
\label{fig:freeze-out}
\end{figure}

As shown in \cref{fig:freeze-out}, along the symmetric thermal-relic estimate the coupling is too small for BSF feedback to generate appreciable elastic $XX$ scattering, although sizable self-interactions may still arise from the baseline $XX^\dagger$ Yukawa interaction for sufficiently small $\mrho$. For the same thermal history, asymmetric DM requires more efficient depletion of the symmetric component and hence $\arho\gtrsim\arho^{\rm sym}(\mX)$, with the primordial $X$--$X^\dagger$ asymmetry adjusted to reproduce the observed abundance~\cite{Baldes:2017gzw}. Increasing $\arho$ suppresses the residual $X^\dagger$ population, and sufficiently large couplings can yield effectively fully asymmetric DM~\cite{Graesser:2011wi,Baldes:2017gzw}. This range overlaps with the parameter space where BSF feedback generates sizable elastic self-interactions, as illustrated in \cref{fig:SIDM_AsymmetricParameterSpace,fig:SIDM_AsymmetricParameterSpace_med_VS_mu,fig:freeze-out}.

The relation between $\arho$ and the relic abundance is cosmology dependent. A different dark-sector temperature at decoupling or subsequent entropy injection would modify the coupling required to reproduce the observed abundance. We therefore treat the self-interaction phenomenology independently of the relic-density requirement, both here and in the companion study~\cite{Petraki:2026SIDMann}.

\subsubsection{Other dark-sector relics}
\label{Sec:SIDM_Cosmology_DarkSectorRelics}

The cosmological fate of the light dark-sector states depends on the realization of the $U(1)_D$ symmetry. We distinguish three possibilities.

\begin{description}

\item[Unbroken $\bm{U(1)_D}$.]
We require $\aV\ll1$ and, in particular, $\aV\ll\arho$, for $V$-mediated interactions to be negligible here. The precise condition depends on screening, e.g. Debye screening in a neutral plasma or stable bound-state formation (cf.~\cref{Sec:SIDM_Cosmology_AsymmDM}). The massless $V$ contributes to dark radiation if cosmologically populated, as does $\varrho$ while relativistic. Planck CMB and BAO measurements constrain additional radiation to approximately $\Delta N_{\rm eff}\lesssim0.5$ at the $3\sigma$ level~\cite{Planck:2018vyg}. This can be accommodated if the dark sector decouples sufficiently early from the SM, so that subsequent entropy release in the visible sector, notably across the QCD transition, lowers the dark-sector temperature relative to the visible bath. The symmetric $\varrho$ population may meanwhile be depleted through $\varrho\varrho^\dagger\to VV$, whose cross section scales as $\aV^2/\mrho^2$ and can remain efficient for sufficiently light $\varrho$ even when $\aV\ll\arho$. Charge neutrality for asymmetric DM is discussed below.

\item[Broken $\bm{U(1)_D}$.]
We assume $V$ to be much heavier than $\varrho$ and sufficiently heavy that vector-mediated interactions are short-ranged and negligible. Decays such as $V\to\varrho\varrho^\dagger$ remove the massive vector relic. Without lighter charged states, an accidental low-energy global symmetry stabilizes $\varrho$. For $\mrho\lesssim{\cal O}({\rm eV})$, a thermal $\varrho$ population can remain an acceptable subdominant hot relic if early decoupling sufficiently suppresses its temperature. For larger $\mrho$, its abundance and conserved charge may instead be transferred to lighter states. For example, fermions $\chi$ with $q_\chi=1$ and $m_\chi<\mrho/2$ may couple through $\varrho^\dagger\overline{\chi^c}\chi+{\rm h.c.}$, analogously to Ref.~\cite{Flores:2025fvd}, allowing the charge-preserving decay $\varrho\to\chi\chi$. This coupling can be weak enough to leave the self-interaction dynamics considered here unaffected. The resulting $\chi$ population is constrained by $N_{\rm eff}$ while relativistic and by relic-abundance and structure-formation considerations if it later becomes nonrelativistic.

\item[Global $\bm{U(1)_D}$.]
If $U(1)_D$ is global from the outset, there is no $V$ relic or associated dark radiation. As in the broken-gauge realization, $\varrho$ may remain as a sufficiently light subdominant relic, or its abundance and charge may be transferred to lighter states while preserving the global symmetry.

\end{description}

\subsubsection{Asymmetric DM and dark charge neutrality}
\label{Sec:SIDM_Cosmology_AsymmDM}

If the gauged $U(1)_D$ is unbroken or mildly broken, an asymmetric $X$ population requires compensating dark charge to maintain gauge neutrality~\cite{Petraki:2014uza}. Minimally, this charge may be carried by $\varrho^\dagger$. Since $\qX=1$ and $\qrho=2$, neutrality requires 
$n_{\varrho^\dagger}-n_\varrho
= (n_X-n_{X^\dagger})/2$. 
Once nonrelativistic, the mass density of this compensating population is suppressed by $\mrho/(2\mX)$ relative to that of $X$ and can therefore be negligible.

Oppositely charged dark-sector species may form Coulombic bound states, potentially modifying the self-interactions~\cite{Petraki:2014uza}. Here, the very small $\aV$ and the potentially small mass of the compensating species imply a low binding scale and tend to suppress dark recombination, suggesting a small composite fraction, though the result depends on the thermal history. If the composite fraction is significant, the resulting multicomponent and composite-DM dynamics would need to be included explicitly in the analysis.

\subsubsection{Inelastic processes inside halos}
\label{Sec:SIDM_Cosmology_RealInelasticity}

The inelastic processes whose feedback on elastic scattering we study here and in the companion paper~\cite{Petraki:2026SIDMann} may also occur as real transitions inside DM halos.

Both $XX^\dagger\to\varrho\varrho^\dagger$ annihilation and radiative BSF can produce indirect-detection signals if a portal to the Standard Model allows the dark-sector radiation to yield visible particles. Bound states formed via $XX\to {\cal B}(XX^\dagger)+\varrho$ are unstable and can subsequently decay into $\varrho\varrho^\dagger$. Long-range interactions have already been shown to enable indirect-detection signals in asymmetric DM, often involving BSF, despite suppressed conventional particle-antiparticle annihilation~\cite{Pearce:2013ola,Pearce:2015zca,Baldes:2017gzw,Baldes:2017gzu,Baldes:2020hwx}. The efficient BSF processes considered here may provide an additional source of such signals. 

If sufficiently efficient, these inelastic processes may deplete DM, potentially excluding parameter regions otherwise compatible with the elastic self-interaction benchmarks. A quantitative assessment lies beyond the scope of the present work and its companion, which focus on the feedback of inelasticity on elastic scattering. Establishing reliable constraints would require a more complete treatment of the inelastic dynamics, including the neglected  dispersive optical-potential contributions from off-shell BSF, as well as the absorptive and dispersive contributions from $\varrho$ bremsstrahlung, along with the evolution of the resulting particle populations. We leave this for future work.
\clearpage
\section{Conclusions}
\label{Sec:Conclusions}

Bound-state formation is a major source of inelasticity in theories with light force mediators. Beyond the direct phenomenological consequences of the inelastic processes themselves, inelasticity necessarily feeds back into elastic scattering, as mandated by the optical theorem. Self-interacting dark matter provides a particularly interesting setting in which this interplay can have important phenomenological consequences. In this work, we have investigated this connection for radiative bound-state formation and studied the elastic scattering generated by its feedback.

We focused on BSF through the emission of a charged boson, for which the interaction governing the incoming scattering state differs from that governing the produced bound state. Such transitions can be exceptionally efficient and arise in both Abelian and non-Abelian theories. In the scenario considered here, BSF proceeds through monopole transitions, which can be even more efficient than the familiar dipole transitions associated with vector emission. These features make this class of processes particularly well suited for generating sizable feedback effects.

We found that BSF necessarily generates an elastic contribution even in the absence of a baseline long-range interaction, as required by unitarity, and that this contribution can become sizable. Its origin differs qualitatively from ordinary long-range elastic scattering: the elastic interaction is generated by an absorptive kernel associated with the inelastic channel, rather than by a soft one-boson-exchange potential acting directly in the incoming state. The long-range dynamics instead enters through the final-state interaction responsible for forming the bound state. Together with the tower of bound levels, this can make capture increasingly efficient toward low velocities and generate a pronounced velocity dependence in the elastic feedback.

We derived analytic estimates for the velocity dependence of BSF-induced elastic scattering and investigated it numerically. For a massless emitted boson, the induced scattering is strongly suppressed at large velocities, but grows rapidly as the velocity decreases. At $\vrel\ll\arho^2$, the total elastic cross section scales as $\sigma_{\reg}^{\elas}\propto\vrel^{-4}$, while the viscosity cross section exhibits a similarly steep velocity dependence. This behavior arises from the increasing number of partial waves contributing to BSF feedback at low velocities, despite the absence of a baseline elastic potential in the incoming state. Unlike Coulomb scattering, however, the partial-wave sums remain convergent at any finite velocity. For a finite mediator mass, the kinematic closure of capture channels into excited bound levels progressively curtails this growth. If ground-state capture remains open at threshold, the induced scattering eventually approaches an $s$-wave plateau; otherwise, it vanishes at zero velocity.  The finite-mediator behavior is illustrated in \cref{fig:SIDM_SymmetricVelocity}. 

The BSF-induced elastic scattering is shaped by principal-value contributions associated with the analytic structure of the inelastic amplitudes, including the tower of bound-state poles. These dispersive effects, which arise from the resummation of the absorptive BSF kernel, can substantially modify the elastic response. Our calculation does not, however, include the direct Hermitian contributions to the optical potential generated by virtual bound-state--mediator intermediate states. Such contributions can persist even when the corresponding BSF channels are kinematically closed, and may further modify both the BSF amplitudes and the induced elastic scattering. Taken together, these features make BSF feedback qualitatively distinct from that of hard inelastic processes such as annihilation, studied in the companion work~\cite{Petraki:2026SIDMann}.

Applied to SIDM, these effects play different phenomenological roles in symmetric and asymmetric scenarios. For symmetric DM, the conventional particle-antiparticle Yukawa interaction typically provides the dominant elastic self-scattering, largely because the finite mediator mass cuts off the low-velocity growth of BSF feedback at higher velocities than that of the Yukawa scattering. The asymmetric scenario instead isolates the more distinctive possibility uncovered in this work: BSF feedback alone can generate elastic self-interactions of phenomenologically relevant magnitude even when the baseline elastic amplitude vanishes.  Depending on the parameters, the resulting velocity dependence can satisfy the cluster constraint while producing sizable scattering at dwarf scales, and also reach the much larger cross sections considered at subhalo scales. This is shown in \cref{fig:SIDM_AsymmetricParameterSpace,fig:SIDM_AsymmetricParameterSpace_med_VS_mu}. Our analysis focuses on the elastic transport generated by the feedback of capture into particle-antiparticle bound states; the effects of real capture and the subsequent evolution of the bound states on the halo composition and energy budget lie beyond the scope of this work.

Several extensions are particularly well motivated. A natural next step is to incorporate the omitted Hermitian dispersive contributions into the BSF feedback calculation. Another direction is to consider BSF feedback in perturbative non-Abelian theories, where capture through the emission of a non-Abelian gauge boson may remain operative without the kinematic cutoff encountered here. More generally, realistic theories may contain multiple light mediators and several important sources of inelasticity acting on the same incoming state. It would therefore be interesting to study the feedback dynamics in this broader setting, allowing several mediators to participate and treating soft or ultrasoft processes such as BSF alongside hard processes such as annihilation, in order to determine how their combined contributions regulate the same elastic scattering amplitude.

\section*{Acknowledgments}
This work was supported by the European Union’s Horizon 2020 research and innovation program under grant agreement No 101002846, ERC CoG CosmoChart. MMF is supported through a FRIPRO grant of the Norwegian Research Council (project ID 353561 ‘DarkTurns’).

\clearpage
\begin{appendices}
\section{Capture via massive-boson emission: bound-level sum}
\label{App:BSFsums}

In this appendix, we derive a smooth analytic approximation for the bound-level sum entering the unregulated BSF cross section, and an exact expression for its leading asymptotic low-velocity coefficient within the Coulomb approximation. These results elucidate the behavior of $r_\ell(\zeta,\gamma)$ and the effect of the finite-mediator phase-space cutoff. In the present model, only even $\ell$ contribute, but we keep $\ell$ general throughout.

\subsection*{General formulas}

For a given incoming partial wave $\ell$, we recall from \cref{eq:BSF_rell_def} that
\begin{align}
r_\ell(\zeta,\gamma) =
\sum_{\ell+1\leqslant n<n_{\rm cut}^{\BSF}(k)}
\Phi_n(k)\,r_{n\ell}(\zeta)
\approx
\int_{\ell+1}^{n_{\rm cut}^{\BSF}(k)}
\dd n\,
\Phi_n(k)\,r_{n\ell}(\zeta) ,
\label{eq:App_BSF_rEllSum}
\end{align}
where the phase-space suppression is defined in \cref{eq:CrossSectionBSF_PhaseSpaceSuppr},
\begin{align}
\Phi_n(k)
=
\sqrt{
1-
\qty[
\frac{\gamma\zeta^2}
{1+\zeta^2/n^2}
]^2
},
\label{eq:App_BSF_Phi_n}
\end{align}
and the strict kinematic cutoff in \cref{eq:BSF_PhaseSpaceCutoff_ncut},
\begin{align}
n_{\rm cut}^{\BSF}(k)
\equiv
\begin{cases}
+\infty,
& \gamma\zeta^2\leqslant1,
\\[1ex]
\dfrac{\zeta}{\sqrt{\gamma\zeta^2-1}},
& \gamma\zeta^2>1.
\end{cases}
\label{eq:App_BSF_nCut}
\end{align}
In the second step of \cref{eq:App_BSF_rEllSum}, the approximation of the discrete sum by the continuum integral assumes that many levels contribute; it becomes less accurate when the kinematic cutoff leaves only a few accessible levels. For the asymptotic low-velocity coefficient, we will return to the exact discrete sum.

The function $r_{n\ell}(\zeta)$ has been defined in \cref{eq:CrossSectionBSF_rnl}. Here, we will use the analytic approximation derived in Ref.~\cite{Flores:2026yay} and employed in Ref.~\cite{Petraki:2025zvv},
\begin{align}
r_{n\ell}(\zeta)
\simeq
2^6\zeta^2
\dfrac{\zeta_n^3}{\qty(1+\zeta_n^2)^3}
\, \abs{
j_\ell\qty(
\frac{2\zeta}{1+\zeta_n^2}
)
}^2,
\qquad
\zeta_n\equiv\frac{\zeta}{n}.
\label{eq:App_BSF_rnlApprox}
\end{align}
This expression is formally obtained for $n\gg\ell$, although it provides a useful approximation to the envelope of the exact result more broadly~\cite{Flores:2026yay,Petraki:2025zvv}. As in the main text, we approximate the attractive bound-state potential by its Coulomb limit, retaining the finite $\mrho$ only in the kinematics of the emitted particle. With
\begin{align}
z
\equiv
\frac{2\zeta}
{1+\zeta^2/n^2},
\label{eq:App_BSF_zDef}
\end{align}
the Jacobian and the factors in \cref{eq:App_BSF_rnlApprox} combine to give
\begin{align}
r_\ell(\zeta,\gamma)
\simeq
8\zeta
\int_{z_{\min}}^{z_{\max}}
\dd z\,
z\,j_\ell^2(z)
\sqrt{
1-
\qty(
\frac{\gamma\zeta z}{2}
)^2
},
\label{eq:App_BSF_rEllIntegral_z}
\end{align}
where
\begin{align}
\label{eqs:App_BSF_zLimits}
z_{\min} \equiv
\dfrac{2\zeta}{1+\zeta^2/(1+\ell)^2}
\qquad \text{and} \qquad
z_{\max} \equiv
\min\qty(2\zeta,\dfrac{2}{\gamma\zeta})
.
\end{align}
The integral has non-empty support only if $z_{\min}<z_{\max}$, equivalently if the lowest level accessible from the $\ell$th partial wave, $n=1+\ell$, satisfies the phase-space condition. For
$\gamma>(1+\ell)^{-2}$, capture from the $\ell$th partial wave therefore shuts off completely at
\begin{align}
\zeta
\geqslant
\frac{1}
{\sqrt{\gamma-(1+\ell)^{-2}}}.
\label{eq:App_BSF_PartialWaveClosure}
\end{align}

\subsection*{Smooth approximation}

To evaluate \cref{eq:App_BSF_rEllIntegral_z} analytically, we use the small- and large-argument forms of the spherical Bessel function~\cite{Flores:2026yay},
\begin{align}
j_\ell(z)
\simeq
\begin{cases}
\dfrac{2^\ell\ell!}{(2\ell+1)!}\,z^\ell,
& 0<z<2b_\ell,
\\[1em]
\dfrac{\sin\qty(z-\ell\pi/2)}{z},
& z\geqslant2b_\ell,
\end{cases}
\label{eq:App_BSF_jEllApprox}
\end{align}
where $b_\ell$ is chosen such that the two approximations for $j_\ell^2$ match upon averaging the rapid oscillations in the second regime,
\begin{align}
\sin^2\qty(z-\ell\pi/2)
\longrightarrow
\frac{1}{2}.
\label{eq:App_BSF_OscAverage}
\end{align}
This gives
\begin{align}
b_\ell
\equiv
\frac{1}{2}
\qty[
\frac{(2\ell+1)!}
{2^{\ell+1/2}\ell!}
]^{1/(\ell+1)}
\simeq
\frac{\ell+1}{e}.
\label{eq:App_BSF_bEll}
\end{align}
The averaging in \cref{eq:App_BSF_OscAverage} removes the rapid oscillations retained in \cref{eq:App_BSF_rnlApprox} and yields a smooth approximation to the envelope of the level sum.\footnote{
This piecewise approximation is useful for estimating the smooth bound-level sum at fixed $\ell$, but is not uniform near the large-order Bessel turning point, as discussed in \cref{Sec:Feedback_VelocityDepend_MasslessW=0}.}

Provided that $z_{\min}<z_{\max}$, three regions arise,
\begin{align}
\begin{array}{llllll}
\text{region A:}
&&
z_{\max}\leqslant 2b_\ell
&\Rightarrow&
\zeta \leqslant b_\ell 
~~\text{or}~~
\zeta\geqslant (\gamma b_\ell)^{-1} 
,
\\[1ex]
\text{region B:}
&&
2b_\ell\leqslant z_{\min}
&\Rightarrow&
\zeta_{-} \leqslant \zeta \leqslant \zeta_{+} 
,
\\[1ex]
\text{region C:}
&&
z_{\min}<2b_\ell<z_{\max}
&\Rightarrow&
b_\ell < \zeta < \zeta_{-}
~~\text{or}~~
\zeta_{+} < \zeta < (\gamma b_\ell)^{-1}
,
\end{array}
\label{eq:App_BSF_Regions}
\end{align}
where
\begin{subequations}
\label{eq:zeta+-}
\label[pluralequation]{eqs:zeta+-}
\begin{gather}
\zeta_{\pm} = 
\dfrac{(\ell+1)^2}{2b_\ell}
\qty(1\pm \sqrt{1-\frac{4b_\ell^2}{(\ell+1)^2}})
\simeq 
(\ell+1) e \qty(\frac{1\pm \sqrt{1-4e^{-2}}}{2}),
\label{eq:zeta+-_def}
\\
\text{or}
\qquad
\zeta_+ \simeq 2.28 \, (1+\ell) 
\qquad \text{and} \qquad
\zeta_- \simeq 0.44 \, (1+\ell) .
\label{eq:zeta+-_num}
\end{gather}
\end{subequations}
It is convenient to introduce
\begin{subequations}
\label{eqs:App_BSF_uDefs}
\begin{align}
u_{\min}
&\equiv
\dfrac{\gamma\zeta}{2} 
\ z_{\min}
=
\frac{\gamma\zeta^2}
{1+\zeta^2/(1+\ell)^2},
\\
u_{\max}
&\equiv
\dfrac{\gamma\zeta}{2} 
\ z_{\max}
=
\min\qty(\gamma\zeta^2,1),
\\
u_b
&\equiv
\gamma\zeta b_\ell.
\end{align}
\end{subequations}
For $\gamma>0$ and within the kinematically allowed region, $0<u_{\min}<u_{\max}\leqslant1$. We further define
\begin{subequations}
\label{eqs:App_BSF_FG}
\begin{align}
F_\ell(u)
&\equiv
{}_2F_1\qty(
-\frac{1}{2},1+\ell;
2+\ell;
u^2
),
\label{eq:App_BSF_Fell}
\\
G(u)
&\equiv
\operatorname{arctanh}\qty(\sqrt{1-u^2})
-
\sqrt{1-u^2}.
\label{eq:App_BSF_G}
\end{align}
\end{subequations}
The resulting smooth approximations are
\begin{subequations}
\label{eq:App_BSF_rellABC}
\label[pluralequation]{eqs:App_BSF_rellABC}
\begin{align}
r_\ell^A(\zeta,\gamma)
&\simeq
\frac{4\zeta}{2\ell+2}
\Bigg[
\qty(
\frac{z_{\max}}{2b_\ell}
)^{2\ell+2}
F_\ell(u_{\max})
-
\qty(
\frac{z_{\min}}{2b_\ell}
)^{2\ell+2}
F_\ell(u_{\min})
\Bigg],
\label{eq:App_BSF_rEllA}
\\[1em]
r_\ell^B(\zeta,\gamma)
&\simeq
4\zeta
\qty[
G(u_{\min})-G(u_{\max})
],
\label{eq:App_BSF_rEllB}
\\[1em]
r_\ell^C(\zeta,\gamma)
&\simeq
4\zeta
\Bigg[
G(u_b)-G(u_{\max})
+
\frac{1}{2\ell+2}
\qty[
F_\ell(u_b)
-
\qty(
\frac{z_{\min}}{2b_\ell}
)^{2\ell+2}
F_\ell(u_{\min})
]
\Bigg].
\label{eq:App_BSF_rEllC}
\end{align}
\end{subequations}
These expressions follow from
\begin{subequations}
\begin{align}
\int \dd u \,
u^{2\ell+1}
\sqrt{1-u^2}
&=
\frac{u^{2\ell+2}}{2\ell+2}
F_\ell(u),
\\
\int \dd u \, \dfrac{\sqrt{1-u^2}}{u}
&= -G(u).
\end{align}
\end{subequations}
\Cref{eqs:App_BSF_rellABC} are Kramers-like relations for individual partial waves describing monopole capture into Coulombic bound states. They generalize the result of Ref.~\cite{Flores:2026yay} to the case of a massive emitted particle.

\subsection*{Massless-emitted-boson limit}

The function $G(u)$ diverges logarithmically as $u\to0$, although the differences appearing in \cref{eqs:App_BSF_rellABC} have finite limits. It is therefore convenient to state the massless case, $\gamma=0$, separately. Then $z_{\max}=2\zeta$, and
\begin{subequations}
\label{eqs:App_BSF_rEllABC_Massless}
\begin{align}
r_\ell^A(\zeta,0)
\simeq{}&
\frac{4\zeta}{2\ell+2}
\qty[
\qty(
\frac{z_{\max}}{2b_\ell}
)^{2\ell+2}
-
\qty(
\frac{z_{\min}}{2b_\ell}
)^{2\ell+2}
],
\label{eq:App_BSF_rEllA_Massless}
\\[1em]
r_\ell^B(\zeta,0)
\simeq{}&
4\zeta
\ln\qty(
\frac{z_{\max}}{z_{\min}}
),
\label{eq:App_BSF_rEllB_Massless}
\\[1em]
r_\ell^C(\zeta,0)
\simeq{}&
4\zeta
\qty[
\ln\qty(
\frac{z_{\max}}{2b_\ell}
)
+
\frac{1}{2\ell+2}
\qty[
1-
\qty(
\frac{z_{\min}}{2b_\ell}
)^{2\ell+2}
]
].
\label{eq:App_BSF_rEllC_Massless}
\end{align}
\end{subequations}
These expressions reproduce the result of Ref.~\cite[Eqs.~(3.15)]{Flores:2026yay}.

\subsection*{Low-velocity limit}

For any fixed nonzero $\gamma$, the asymptotic low-velocity behavior differs qualitatively from the massless-emission limit. We first consider
$0<\gamma<(1+\ell)^{-2}$, for which at least one bound level remains kinematically accessible as $\zeta\to\infty$. In this limit,
\begin{align}
z_{\min}\simeq\frac{2(1+\ell)^2}{\zeta},
\qquad
z_{\max}\simeq\frac{2}{\gamma\zeta},
\end{align}
so the smooth approximation eventually lies entirely in region A. To determine not only the velocity dependence but also its coefficient, we return to the discrete sum (the first equality) in \cref{eq:App_BSF_rEllSum}, rather than approximating it by an integral.

For each fixed bound level $n$, the exact Coulombic expression in \cref{eq:CrossSectionBSF_rnl} gives, at $\zeta\to\infty$,
\begin{align}
r_{n\ell}(\zeta)
&\simeq
\qty[\frac{2^{2\ell+3}\ell!}{(2\ell+1)!}]^2
\frac{n^{2\ell+2}(n+\ell)!}{(n-\ell-1)!}
\frac{1}{\zeta^{2\ell+1}},
\label{eq:App_BSF_rnl_LargeZeta}
\end{align}
while, for $\gamma n^2<1$, the phase-space factor satisfies
\begin{align}
\Phi_n(k)\longrightarrow\sqrt{1-\gamma^2n^4}.
\end{align}
Since the number of accessible levels approaches a finite value at fixed $\gamma>0$, the leading asymptotic behavior can be summed term by term:
\begin{subequations}
\label{eqs:App_BSF_LargeZeta}
\begin{align}
r_\ell(\zeta,\gamma)
&\simeq
\frac{A_\ell(\gamma)}{\zeta^{2\ell+1}},
\qquad
\zeta\to\infty,
\quad 0<\gamma<(1+\ell)^{-2},
\label{eq:App_BSF_LargeZetaScaling}
\\[0.5em]
A_\ell(\gamma)
&\equiv
\qty[\frac{2^{2\ell+3}\ell!}{(2\ell+1)!}]^2
\sum_{\ell+1\leqslant n<\gamma^{-1/2}}
\frac{n^{2\ell+2}(n+\ell)!}{(n-\ell-1)!}
\sqrt{1-\gamma^2n^4}.
\label{eq:App_BSF_LargeZetaCoefficient}
\end{align}
\end{subequations}
The coefficient $A_\ell(\gamma)$ is exact at leading order in $1/\zeta$ within the Coulomb and phase-space approximations employed here, even when only a few bound levels remain accessible. The $\gamma\to0$ and $\zeta\to\infty$ limits do not commute; in particular, the above expansion should not be applied at $\gamma=0$.

For fixed $\ell$ and $\gamma\to0$, many levels contribute, with the dominant terms occurring at $n\sim\gamma^{-1/2}\gg\ell$. Using $(n+\ell)!/(n-\ell-1)!\simeq n^{2\ell+1}$ in \cref{eq:App_BSF_LargeZetaCoefficient} and replacing the sum by an integral, we obtain
\begin{align}
A_\ell(\gamma)
&\simeq
\qty[\frac{2^{2\ell+3}\ell!}{(2\ell+1)!}]^2
\frac{B\qty(\ell+1,\frac32)}{4\gamma^{2\ell+2}},
\qquad
\gamma(1+\ell)^2\ll1,
\label{eq:App_BSF_LargeZetaCoefficient_SmallGamma}
\end{align}
where $B$ is the Euler beta function. This approximation is not uniform at large $\ell$. In particular,
\begin{align}
A_0(\gamma)
&=64\sum_{1\leqslant n<\gamma^{-1/2}}
n^3\sqrt{1-\gamma^2n^4}
\simeq\frac{32}{3\gamma^2},
\qquad \gamma\ll1,
\label{eq:App_BSF_LargeZetaCoefficient_sWave}
\end{align}
which is the coefficient used for the $s$-wave plateau in \cref{Sec:Feedback_VelocityDepend_Massive}.

Thus, at fixed $\gamma>0$ with open capture channels, the finite mediator mass replaces the massless-emission enhancement by the threshold scaling $\sigma_{\ell,\unreg}^{\BSF}\vrel\propto\vrel^{2\ell}$ and suppresses higher partial waves at very low velocities. If $\gamma>(1+\ell)^{-2}$, capture from the $\ell$th partial wave instead closes completely at the finite value of $\zeta$ given in \cref{eq:App_BSF_PartialWaveClosure}. At the boundary $\gamma=(1+\ell)^{-2}$, the $n=1+\ell$ level remains accessible for every finite $\zeta$, while higher levels eventually close; its phase-space factor approaches $\Phi_{\ell+1}(k)\simeq\sqrt{2}(\ell+1)/\zeta$, yielding the stronger suppression
\begin{align}
r_\ell\qty(\zeta,(1+\ell)^{-2})
&\simeq
\frac{2^{4\ell+6}\sqrt{2}\,(\ell!)^2\,(1+\ell)^{2\ell+3}}
{(2\ell+1)!\,\zeta^{2\ell+2}}.
\label{eq:App_BSF_LargeZeta_Threshold}
\end{align}

\clearpage
\end{appendices}

\newpage
\phantomsection
\addcontentsline{toc}{section}{References} 
\bibliography{biblio.bib}
\bibliographystyle{JHEP.bst}
\end{document}